\documentclass[aps,pra,reprintnumbers,floatfix]{revtex4}
\usepackage{epsfig}
\usepackage{graphicx}
\usepackage{dcolumn}
\usepackage{amsmath,amssymb}
\usepackage{pstricks}
\usepackage{color}
\usepackage{footnote}
\usepackage{bm}
\usepackage{bbold}
\usepackage[colorlinks=true,citecolor={blue},urlcolor={red},linkcolor={blue}]{hyperref}

\begin{document}


\title{Study of laser-driven multielectron dynamics of Ne atom using time-dependent optimized second-order many-body perturbation theory}
\author{Himadri Pathak,$^{1,}$\footnote{\color{blue}pathak@atto.t.u-tokyo.ac.jp}
Takeshi Sato,$^{1,2,3,}$\footnote{\color{blue} sato@atto.t.u-tokyo.ac.jp}
and Kenichi L. Ishikawa,$^{1,2,3}$\footnote{\color{blue}ishiken@n.t.u-tokyo.ac.jp}}
\affiliation{$^1$Department of Nuclear Engineering and Management, School of Engineering, The University of Tokyo, 7-3-1 Hongo, Bunkyo-ku, Tokyo 113-8656, Japan}
\affiliation{$^2$Photon Science Center, School of Engineering, The University of Tokyo, 7-3-1 Hongo, Bunkyo-ku, Tokyo 113-8656, Japan}
\affiliation{$^3$Research Institute for Photon Science and Laser Technology, The University of Tokyo, 7-3-1 Hongo, Bunkyo-ku, Tokyo 113-0033, Japan}
\begin{abstract}
We calculate the high-harmonic generation (HHG) spectra, strong-field ionization, and time-dependent dipole-moment
of Ne using explicitly time-dependent optimized
second-order many-body perturbation method (TD-OMP2) where both orbitals and amplitudes are time-dependent.
We consider near-infrared (800 nm) and mid-infrared (1200 nm) laser pulses with very high intensities (5$\times$10$^{14}$,
8$\times$10$^{14}$, and 1$\times$10$^{15}$ W/cm$^2$), required for strong-field experiments with the high-ionization potential (21.6 eV) atom.
We compare the result of the TD-OMP2 method with the time-dependent complete-active-space
self-consistent field method and the time-dependent Hartree-Fock method.
Further, we report the implementation of the TD-CC2 method within the chosen active space, which is also a second-order approximation to the TD-CCSD method,
and present results of time-dependent dipole-moment and HHG spectra 
with an intensity of 5$\times$10$^{13}$ W/cm$^2$ at a wavelength of 800 nm.
It is found that the TD-CC2 method is not stable in the case with a higher laser intensity,
and it does not provide a gauge-invariant description of the physical properties, which makes TD-OMP2 a superior choice to reach out to larger chemical systems,
especially for the study of strong-field dynamics.
The obtained results indicate that the TD-OMP2 method shows moderate performance, overestimating the response of Ne, while TDHF underestimates it.
Nevertheless, it is remarkable that stable computation of such highly nonlinear nonperturbative phenomena is possible
within the framework of time-dependent perturbation method, by virtue of the nonperturbative inclusion of the laser-electron
interaction and time-dependent optimization of orbitals.
\end{abstract}
\maketitle

\section{Introduction}\label{intro}
There has been increasing interest in the atomic, molecular, and solid-state response to an ultrashort intense laser pulse
\cite{paul2001observation, hentschel2001attosecond, schultze2010, klunder2011probing, belshaw2012observation, calegari2014, itatani2004, smirnova2009high, haessler2010attosecond}.
The recent advances in laser technology have made it possible to observe electron dynamics on attosecond time scales, opening up possibilities of new spectroscopic and measurement 
methods \cite{corkum2007attosecond, krausz2009attosecond, itatani2004, baker2006probing, goulielmakis2010real, sansone2010electron}.
These new techniques will eventually lead to understanding yet unexplored pertinent areas of research with unprecedented  time resolution.\par
High-harmonic generation (HHG), in which a fundamental strong laser field is converted into harmonics of very high orders, is an avenue to generate coherent attosecond light pulses in the spectral range from extreme-ultraviolet (XUV) to the soft X-ray regions \cite{antoine1996attosecond}.
HHG process is by nature highly non-linear, and its spectrum has a distinctive shape; a plateau, where the intensity of the emitted radiation 
remains nearly constant up to many orders, and then an abrupt cutoff, beyond which practically no harmonics are observed \cite{kli2010}. 
These features can be intuitively explained by the semiclassical three-step model \cite{corkum1994plasma, kulander1993super};
(i) an electron escapes to the continuum at the nuclear position with zero kinetic energy through tunnel ionization, (ii) it moves classically and is driven back toward the parent ion by the laser field, and (iii) when the electron comes back to the parent ion and possibly recombines with it, a harmonic photon whose energy is the sum of the electron kinetic energy and the ionization potential $I_p$ is emitted. 
Then, the cutoff energy $E_c$ is given by $E_c=I_p+3.17U_p$, where $U_p=E_0^2/4\omega_0^2$ denotes the ponderomotive energy ($E_0$: laser electric field strength, $\omega_0$: carrier frequency).

The time-dependent Schr{\"o}dinger equation (TDSE) provides the rigorous theoretical description of the laser-induced multielectron dynamics.
However, direct real-space solutions of the TDSE beyond two-electron systems remains a major challenge \cite{parker1998intense, parker2000time, pindzola1998time, laulan2003correlation, ishikawa2005above, feist2009probing, ishikawa2012competition, sukiasyan2012attosecond, vanroose2006double, horner2008classical}.
As a consequence, the single-active-electron (SAE) approximation \cite{krause1992jl, kulander1987time} is widely used, in which only the outermost electron is explicitly treated.
Laser-induced multielectron dynamics is, however, beyond the reach of this approximation.

\begin{figure}[hb!]
\centering
\begin{center}
\includegraphics[width=0.8\linewidth]{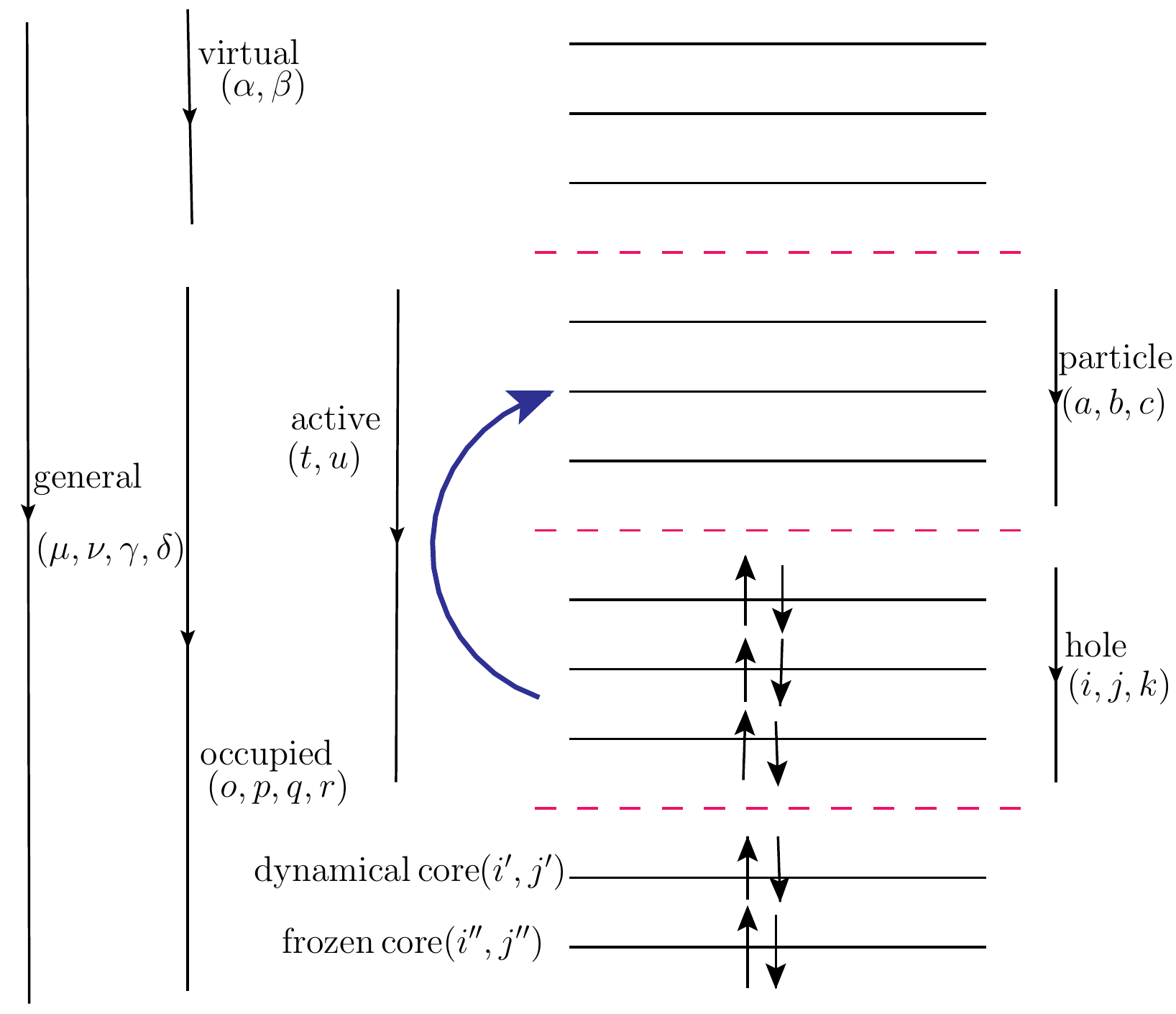}
\caption{The orbital sub-spacing for a spin-restricted case.
The horizontal lines represent spatial orbitals, divided into frozen-core, dynamical core, and active. 
The active orbital space is further split into the hole and particle subspaces those occupied and virtual with respect to the Hartree-Fock determinant. 
The up and down arrows represent electrons.}
\label{subspace}
\end{center}  
\end{figure}

One of the most advanced methods to describe the multielectron dynamics
is the multiconfiguration time-dependent
Hartree-Fock (MCTDHF) method \cite{caillat2005correlated, kato2004time,
nest2005multiconfiguration, haxton2011multiconfiguration,
hochstuhl2011two, RevModPhys.92.011001} and more generally the 
time-dependent multiconfiguration self-consistent-field (TD-MCSCF) method.
In TD-MCSCF, the electronic wavefunction is given by the
configuration-interaction (CI) expansion, and both CI coefficients
and spin-orbital functions constituting the Slater determinants are
propagated in time.
The applicability of the full CI-based MCTDHF method is significantly
broadened by the time-dependent complete-active-space
self-consistent-field (TD-CASSCF) method \cite{sato2013time}, which
introduces the frozen-core, dynamical-core, and active orbital
subspaces as illustrated in Fig. \ref{subspace}.
More approximate, and thus computationally more efficient
methods \cite{miyagi2013time, miyagi2014time,
haxton2015two,sato2015time} have been developed by relying on a truncated CI
expansion within the active orbital space, compromising the size
extensivity condition. 

To restore the size-extensivity, the choice of the coupled-cluster expansion
\cite{shavitt:2009, kummel:2003, crawford2007introduction} within the
time-dependent active orbitals is a worthy one. 
This idea was first realized by the orbital-adapted time-dependent
coupled-cluster (OATDCC) method \cite{kvaal2012ab}, which is based on
the complex analytic action functional using the biorthonormal orbitals. 
We have also developed time-dependent optimized coupled-cluster (TD-OCC) method
\cite{sato2018communication}, based on the real action functional using
time-dependent orthonormal orbitals.

Recently, to further extend the applicability to heavier atoms and larger
molecules interacting with intense laser fields, we have implemented
approximate methods within the TD-OCC framework without losing the
size-extensivity criteria \cite{pathak2020time,pathak2020mp2}. 
In Ref.~\cite{pathak2020time}, we introduced a method designated as
TD-OCEPA0, which is based on the simplest version of the
coupled-electron pair approximation \cite{bozkaya2011quadratically}.
We have also implemented an approximate method in the TD-OCC framework,
based on the many-body perturbation expansion of the coupled-cluster effective Hamiltonian \cite{pathak2020mp2}.
This method, designated as time-dependent orbital-optimized second-order
many-body perturbation method (TD-OMP2), is a time-dependent extension
of the orbital-optimized MP2 method developed by Bozkaya
{\it et al,} \cite{bozkaya2011quadratically} for the stationary
electronic structure calculations. It is size-extensive, 
gauge-invariant, and has a lower scaling of the
computational cost [$O(N^5)$ where $N$ is the number of active orbitals]
than the TD-OCC method with double excitations (TD-OCCD) having a scaling of $O(N^6)$.
For both the TD-OMP2 and TD-OCEPA0 \cite{pathak2020time,pathak2020mp2}, 
one need not solve for de-excitation
amplitudes since they are the complex conjugate of the excitation
amplitudes, which further leads to a significant reduction in the computational cost.

Furthermore, in the present work, we report the implementation of the so-called CC2 method \cite{christiansen1995second} within the  active space in
the time-dependent framewor (TD-CC2).
The CC2 method is also a second-order approximation to the coupled-cluster singles and doubles (CCSD) model.
In this method, the doubles equation is approximated to provide the first-order corrections to the wavefunction, 
and the singles equation is kept the same as in the CCSD approximation.
It scales $N^5$ and produces ground state comparable to the MP2 method.
The equations of motions (EOMs) are derived based on the real-action formulation with orthonormal orbital functions, following our earlier work \cite{sato2018communication}.
We have omitted hole-particle rotations in our implementation
while retaining the single
amplitudes \cite{scuseria1987optimization}. 
The major drawback of the TD-CC2 method is the lack of gauge invariance, as numerically shown in this work. 

There are numerous experiments performed on the noble gas atoms and their mixtures
\cite{PhysRevLett.79.2967, PhysRevLett.70.774, PhysRevLett.70.766,
PhysRevLett.99.053904, PhysRevA.66.021802, takahashi2013attosecond}.
Harmonics of higher than 300 orders have been obtained.
The TDDFT \cite{PhysRevLett.74.872} is an attractive choice to study strong-field phenomena for larger chemical systems {\cite{PhysRevA.101.063413}.
Shih-I Chu {\it et al.} \cite{PhysRevA.57.452} developed the self-interaction-free time-dependent density-functional theory (TDDFT)
and extensively studied laser-driven dynamics in noble gas atoms \cite{PhysRevA.64.013417}, and also in heteronuclear diatomics \cite{PhysRevA.83.043414}.
However, their method \cite{PhysRevA.57.452} is not free from the general drawbacks of the DFT.
The TD-OMP2 is a choice in the wavefunction based methods to reach out to larger chemical systems studying of strong-field dynamics with affordable scaling.}

In this article, we apply the TD-OMP2 method to the study of
laser-induced dynamics in Ne atom.
Ne is having the highest ionization potential value among the noble gas atoms (except for the He, for which anyway it is possible to have an exact solution of the 
TDSE \cite{PhysRevLett.107.093005,PhysRevA.72.013407,PhysRevLett.108.033003}), and 2$s$ and 2$p$ orbitals are well separated from each other.
These make Ne as an interesting candidate to deal with as a test case.
It is also true that highly accurate results can be produced by the method like TD-CASSCF to have a
better understanding of the capability of the newly implemented approximate methods before moving to larger chemical systems. 

First, we seek for the most suitable
active space configuration and the maximum angular momentum to expand
the time-dependent orbitals for the study with the highest employed intensity
by performing a series of calculations.  Then, we compare the TD-OMP2
results with those of time-dependent Hartree-Fock (TDHF) and the TD-CASSCF for the highest employed intensity.
Further, we report the results of time-dependent dipole-moment calculated using the TD-CC2 method and compare it with other methods
and demonstrate that the property evaluated using this method is not gauge-invariant.
We also report a comparison of the computational timing between the TD-OMP2 and TD-CC2 methods.
The manuscript is organized as follows. A concise description of the TD-OMP2 and TD-CC2 method is presented in Sec. \ref{sec2}.
Section~\ref{sec3} reports and discusses the numerical results.
Finally, concluding remarks are given in Sec. \ref{sec4}.
We use Hartree atomic units unless stated otherwise, and Einstein convention is implied
for summation over orbital indices.

\section{Method} \label{sec2}
We consider a system with $N$ electrons governed by the following
time-dependent Hamiltonian,
\begin{eqnarray}\label{eq:ham1q}
H(t) &=& \sum_{i=1}^N
h(\bm{r}_i,\bm{p}_i,t)
+ \sum_{i=1}^{N-1}\sum_{j=2}^N \frac{1}{|\bm{r}_i-\bm{r}_j|},
\end{eqnarray}
where $\bm{r}_i$ and $\bm{p}_i$ are the position and canonical momentum
of an electron {\color{black}$i$}. The corresponding second quantized Hamiltonian reads
\begin{eqnarray}\label{eq:ham2q}
\hat{H}
&=& h^\mu_\nu \hat{c}^\dagger_\mu\hat{c}_\nu +
 \frac{1}{2}u^{\mu\gamma}_{\nu\lambda} \hat{c}^\dagger_\mu\hat{c}^\dagger_\gamma\hat{c}_\lambda\hat{c}_\nu,
\end{eqnarray}
where $\hat{c}^\dagger_\mu$ ($\hat{c}_\mu$) is a creation
(annihilation) operator for the complete, orthonormal set of spin-orbitals $\{\psi_\mu(t)\}$, which are explicitly time-dependent, and
\begin{eqnarray}
h^\mu_\nu = \int dx_1 \psi^*_\mu(x_1) h(\bm{r}_1,\bm{p}_1) \psi_\nu(x_1),
\end{eqnarray}
\begin{eqnarray}
 u^{\mu\gamma}_{\nu\lambda} = \int\int dx_1dx_2
 \frac{\psi^*_\mu(x_1)\psi^*_\gamma(x_2)\psi_\nu(x_1)\psi_\lambda(x_2)}{|\bm{r}_1-\bm{r}_2|},
\end{eqnarray}
where $x_i=(\bm{r}_i,\sigma_i)$ is a composite spatial-spin coordinate.
Hereafter we refer to spin-orbitals simply as orbitals, and
use orbital indices $i,j,k\cdots$ to denote orbitals in the hole space
which are occupied in a reference determinant $\Phi$, and
$a,b,c,\cdots$ for those in the particle space which are unoccupied in
the reference and accommodate excited 
electrons. We use $p,q,r,\dots$ for general active orbitals (union of hole
and particle). It should be noted that, as an orbital-optimized theory,
the number of active orbitals is less than the number of the full set
of the orbitals $\{\psi_\mu\}$ in general \cite{sato2013time,sato2015time,sato2018communication,pathak2020time,pathak2020mp2}.

\begin{figure}[hb!]
\centering
\begin{center}
\includegraphics[width=1.0\linewidth]{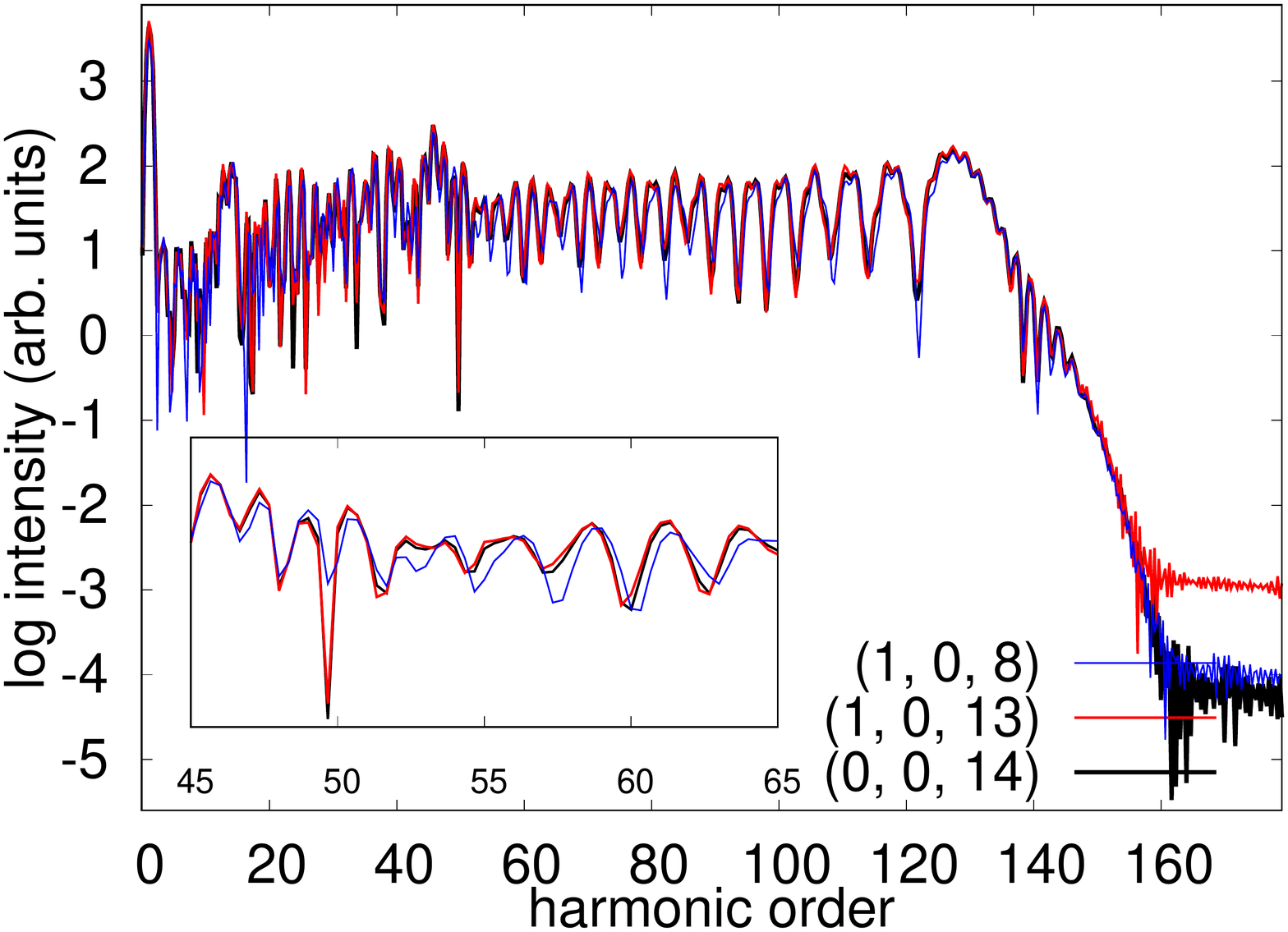}
\caption{HHG spectra of Ne exposed to a laser pulse with a wavelength of 800 nm and an intensity of 1$\times$10$^{15}$ W/cm$^2$.
Results of the TD-OMP2 method with different numbers of orbital configuration (m, n, o) and maximum angular momentum $L_{max}$=63.}
\label{hhg_active_space}
\end{center}  
\end{figure}

\subsection{Review of TD-OMP2 method}
The ground-state MP2 method can be viewed as an approximation to the CCD method considering only those terms which give first-order contributions
to the wavefunction with respect to the fluctuation potential \cite{helgaker2014molecular}.
To construct the TD-OMP2 method as a time-dependent, orbital-optimized
counterpart of the MP2 method, we begin with the
time-dependent CCD Lagrangian, and
retain only those terms giving up to second-order contributions to the 
Lagrangian \cite{pathak2020mp2}, which reads
\begin{eqnarray}\label{eq:td-omp2_lag}
L &=&  L_0 -i\lambda^{ij}_{ab}\dot{\tau}^{ab}_{ij} +
\langle\Phi|\hat{\Lambda}_2(\hat{H}-i\hat{X})|\Phi\rangle \\
&+&
\langle\Phi|[\hat{H}-i\hat{X},\hat{T}_2]|\Phi\rangle +
\langle\Phi|\hat{\Lambda}_2[\hat{f}-i\hat{X},\hat{T}_2]|\Phi\rangle, \nonumber
\end{eqnarray}
where $L_0 = \langle\Phi|(\hat{H}-i\partial/\partial t)|\Phi\rangle$,
$\hat{f}=f^\mu_\nu \hat{c}^\dagger_{\mu} \hat{c}_\nu$, 
$\hat{X}=X^\mu_\nu \hat{c}^\dagger_{\mu} \hat{c}_\nu$,
$f^\mu_\nu \equiv h^\mu_\nu + v^{\mu j}_{\nu j}$, 
$X_\mu^\nu=\langle \psi_\nu|\dot \psi_\mu\rangle$,
$v^{pr}_{qs}\equiv
u^{pr}_{qs}-u^{pr}_{sq}$, $\hat{T}_2=\tau^{ab}_{ij}\hat c_a^\dag \hat c_b^\dag \hat c_j \hat c_i$, and
$\hat{\Lambda}_2=\lambda^{ij}_{ab}\hat c_i^\dag \hat c_j^\dag \hat c_b \hat c_a$.
Then the action functional
\begin{eqnarray}
S &=& \label{eq:action}
\int_{t_0}^{t_1} L(t)dt
\end{eqnarray}
is required to be stationary, $\delta S = 0$, with respect to the
variation of the amplitudes $\{\tau^{ab}_{ij}\}$, $\{\lambda^{ij}_{ab}\}$ and
orthonormality-conserving variation of orbitals
\cite{pathak2020mp2}. The resultant EOM for amplitudes reads
\begin{eqnarray}
i\dot{\tau}^{ab}_{ij}=\label{t2eqn}
v_{ij}^{ab}-p(ij) f_j^k
 \tau_{ik}^{ab}+p(ab)f_c^a \tau_{ij}^{cb}
, \hspace{2em} 
\lambda^{ij}_{ab}=\tau^{ab*}_{ij}
\end{eqnarray}
where $p(ij)$ and $p(ab)$ are the cyclic permutation operator.
We have also arbitrarily chosen one of the orbital
gauges $\langle\psi_i|\dot{\psi}_j\rangle=\langle\psi_a|\dot{\psi}_b\rangle=0$, using the
invariance of the total wavefunction with respect to the unitary
transformation within the hole and particle spaces separately to simplify the equations of motion.

The EOMs for orbitals are given by
\begin{eqnarray}
i|\dot{\psi_p}\rangle = \label{eq:eom_orb}
(1-\hat{P}) \hat{F}_p|\psi_p\rangle + |\psi_q\rangle X^q_p,
\end{eqnarray}
\begin{eqnarray}
&i\left\{(\delta^a_b\rho^j_i-\rho^a_b\delta^j_i)X^b_j\right\}= F^a_j\rho^j_i - \rho^a_bF^{i*}_b
\end{eqnarray}
\begin{eqnarray}
\hat{F}_p|\psi_p\rangle = \label{eq:eom_gfockoperator}
\hat{h} |\psi_p\rangle +  \hat{W}^r_s|\psi_q\rangle \rho^{qs}_{or}{(\rho^{-1})}_p^o,
\end{eqnarray}
\begin{eqnarray}
W^r_s(\bm{x}_1)=
\int d\bm{x}_2
\frac{\psi^*_r(\bm{x}_2)\psi_s(\bm{x}_2)}{|\bm{r}_1-\bm{r}_2|}, 
\end{eqnarray}
where $\hat{P}=\sum_p|\psi_p\rangle\langle\psi_p|$, $F_q^p=\langle
\phi_p|F_q|\phi_q\rangle$, 
and $\rho_q^p$ and $\rho_{qs}^{pr}$ are the one- and two-body reduced
density matrices given by
\begin{eqnarray}
\rho^q_p = \gamma^q_p + \delta^q_j\delta^j_p, \rho^{qs}_{pr} = \Gamma^{qs}_{pr} +
\gamma^q_p \delta^s_j \delta^j_r
+\gamma^s_r \delta^q_j\delta^j_p
-\gamma^q_r \delta^s_j\delta^j_p
-\gamma^s_p \delta^q_j\delta^j_r
+\delta^q_j\delta^j_p\delta^s_k\delta^k_r
-\delta^s_j\delta^j_p\delta^q_k\delta^k_r,\,\,\,\,\,\,\,\,\,\,\,\,\,\,\,\
\end{eqnarray}
with non-zero elements of $\gamma^q_p$ and $\Gamma^{qs}_{pr}$ being
\begin{eqnarray}
\gamma_j^i=-\frac{1}{2}{\tau_{ki}^{cb}}^{\dag}\tau_{kj}^{cb},
\gamma_a^b=\frac{1}{2}{\tau_{kl}^{ca}}^{\dag}\tau_{kl}^{cb},
\Gamma_{ij}^{ab}=\tau_{ij}^{ab}, \Gamma_{ab}^{ij}={\tau_{ij}^{ab}}^{\dag}.
\end{eqnarray}
\subsection{TD-CC2 method}
The stationary CC2 method can be viewed as an approximation to the CCSD method considering only those terms which give first-order contributions
to the wavefunction with respect to the fluctuation potential \cite{christiansen1995second}.
To construct the TD-CC2 method as a time-dependent counterpart of the CC2 method, we shall begin with the time-dependent CCSD Lagrangian, 
and retain only those terms giving up to second-order contributions to the wavefunction, which reads
\begin{eqnarray}\label{eq:td-cc2_lag}
L &=&  L_0 -i\lambda^{i}_{a}\dot{\tau}^{a}_{i} -i\lambda^{ij}_{ab}\dot{\tau}^{ab}_{ij} + 
\langle\Phi|(\hat{\Lambda}_1+\hat{\Lambda}_2)(\bar{H}-i\bar{X})|\Phi\rangle \\
&+&
\langle\Phi|[\bar{H}-i\bar{X},\hat{T}_2]|\Phi\rangle +
\langle\Phi|\hat{\Lambda}_1[\bar{H}-i\bar{X},\hat{T}_2]|\Phi\rangle
 +
\langle\Phi|\hat{\Lambda}_2[\bar{f}-i\bar{X},\hat{T}_2]|\Phi\rangle,\nonumber
\end{eqnarray}
where $\bar O\equiv e^{-\hat T_1} \hat{O}e^{\hat T_1}$, $\hat{T}_1 = \tau^a_i \hat{c}^\dagger_a \hat{c}_i$, and
$\hat{\Lambda}_1 = \lambda^i_a \hat{c}^\dagger_i \hat{c}_a$.
It should be noticed that the singles amplitudes $\tau^a_i, \lambda^i_a$ are treated as a zeroth-order quantity.
Then, following the real-valued action formulation as described in the previous section, we derive the amplitude EOMs as 
\begin{eqnarray}
&&i\dot \tau_i^a=\langle \Phi_i^a|\bar H-i\bar {X}+[\bar H-i\bar{X}, \hat T_2]|\Phi\rangle,\label{t1cc2eqn}
\end{eqnarray}
\begin{eqnarray}
i\dot \tau_{ij}^{ab}=\langle \Phi_{ij}^{ab}|\bar H-i\bar{X}+[\bar f-i\bar{X}, \hat T_2]|\Phi\rangle\label{t2cc2eqn}, 
\end{eqnarray}
\begin{eqnarray}
&&-i\dot \lambda_a^i=\langle\Phi|(1+\hat \Lambda_1)[\hat H-i\hat {X}+[\hat H-i\hat {X}, \hat T_2],\hat E_{i}^{a}]|\Phi\rangle\nonumber\\
&&\hspace{2cm}+\langle\Phi|\hat \Lambda_2[\hat H-i\hat {X}+[\hat f-i\hat {X}, \hat T_2],\hat E_{i}^{a}]|\Phi\rangle, \label{l1cc2}\\ 
&&-i\dot \lambda_{ab}^{ij}=\langle\Phi|(1+\hat \Lambda_1)[\bar H -i\bar {X},\hat E_{ij}^{ab}]|\Phi\rangle+\langle\Phi|\hat \Lambda_2[\bar f-i\bar {X}, \hat E_{ij}^{ab}]|\Phi\rangle\label{l2cc2},
\end{eqnarray}
Expanding right-hand sides of the Eq.~\ref{t1cc2eqn}, \ref{t2cc2eqn}, \ref{l1cc2}, and \ref{l2cc2} in terms of one-body and two-body matrix elements obtain programmable algebraic 
expressions, Eq.~\ref{t1matrix}, \ref{t2matrix}, \ref{l1matrix}, and \ref{l2matrix}.

The EOMs for orbitals are derived as
\begin{eqnarray}
i|\dot{\psi_p}\rangle = \label{eq:eom_orb_cc2}
(1-\hat{P}) \hat{F}_p|\psi_p\rangle, 
\end{eqnarray}
which is formally identical to those for TD-OMP2, Eq.~\ref{eq:eom_orb}, except for the absence of the second term, where 
(i) again we arbitrarily chose X$_{i}^{j}$ = X$_{a}^{b} = 0$, and (ii) the hole-particle rotations are also fixed as $X_{i}^{a} = 0$). We adopted the latter, because both real- and imaginary-time propagation encounters convergence difficulty due to similar roles played by $\hat{T}_1$ amplitudes and hole-particle rotations. (See \cite{scuseria1987optimization, sato2018communication} for the related discussions for the stationary problems.)
The correlation contributions of the density matrices for the TD-CC2 method are given in Eq.~\ref{cc2_1rdm} and \ref{cc2_2rdm}.


In summary, TD-CC2 method differs from the TD-OMP2 method in that it 
includes the singles amplitudes and ignores the hole-particle rotations.
While the former treatment (CC2) is usually preferred in the stationary theory, we consider that the latter approach (TD-OMP2) is advantageous in the time-dependent problems, since the omission of the hole-particle rotation results in the loss of gauge-invariance of the TD-CC2 method, as numerically demonstrated in the next section.

\section{Results and discussion: Application to electron dynamics in Ne}\label{sec3}

In this section, we report and discuss the application of our numerical
implementation of the TD-OMP2 method to laser-driven electron dynamics
in Ne atom. Within the dipole approximation in the velocity gauge,
the one-electron Hamiltonian is given by
\begin{eqnarray}
h(\bm{r},\bm{p}) = \frac{1}{2}|\bm{p}^2| - \frac{Z}{|\bm{r}|} + A(t)p_z
\end{eqnarray}
where $Z=10$ is the atomic number, $A(t) = -\int^t E(t^\prime)dt^\prime$ is the vector potential,
with $E(t)$ being the laser electric field linearly polarized along $z$
axis, given as
\begin{figure}[hb!]
\centering
\begin{center}
\includegraphics[width=1.1\linewidth]{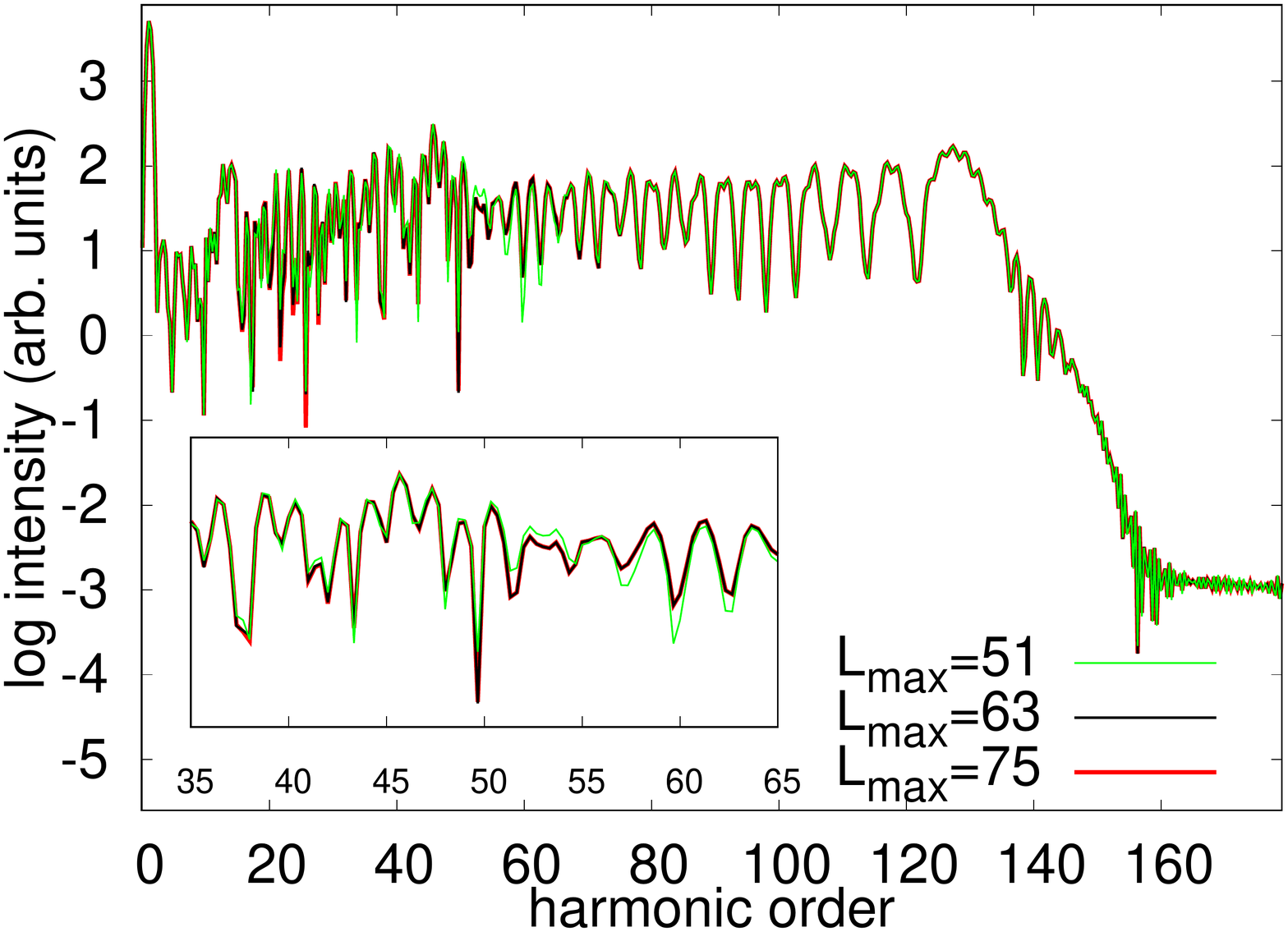}
\caption{HHG spectra of Ne exposed to a laser pulse with a wavelength of 800 nm and an intensity of 1$\times$10$^{15}$ W/cm$^2$.
Results of the TD-OMP2 method obtained with different maximum angular
 momentum L$_{max}$ with the orbital configuration (1,0,13).}
\label{hhg_lmax}
\end{center}  
\end{figure}
\begin{eqnarray}
E(t)=E_0\,{{\sin}}(\omega_0t)\,{{\sin}}^2\left(\pi\frac{t}{\tau}\right),\,\,\,\, 0\le t \le \tau,
\end{eqnarray}
with a foot-to-foot pulse duration $\tau$ (3T), a peak intensity $I_0=E_0^2$,
period $T=2\pi/\omega_0$, and a wavelength of $\lambda=2\pi/\omega_0$. 
In our implementation, the time-dependent orbitals are expanded with spherical-FEDVR basis functions,
\begin{eqnarray}
\chi_{klm}(r, \theta, \psi)=\frac{1}{r}f_k(r)Y_{lm}(\theta, \phi) 
\end{eqnarray}
where, $Y_{lm}$ and $f_k(r)$ are spherical harmonics and the normalized radial-FEDVR basis function \cite{fedvr1, fedvr2}, respectively.
The spherical harmonics expansion is continued up to the maximum angular
momentum of $L_{max}$, and the radial FEDVR basis supports the range of
radial coordinate $0\leq r \leq R_{max}$, with an appropriate absorbing
boundary condition. 
The details of the implementation can be found in \cite{sato2016time, orimo2018implementation}.
We have used fourth-order exponential Runge-Kutta integrator \cite{hochbruck2010exponential} to propagate equations of motions
with 10000 time-steps per optical cycle. The simulations are run for further 3000 time steps after the end of the pulse.
We have used a regularization while inverting the one-body reduced density matrix in Eq. (\ref{eq:eom_gfockoperator}).
Details of the functional form of the regularizer can be found in ref \cite{sato2015time}.

First, we seek for the optimum orbital space for our study,
by changing the numbers ($n_{\rm fc}$,$n_{\rm dc}$,$n_{\rm
act}$), where $n_{\rm fc}$ is the number of frozen-core orbitals which
are forced to be doubly occupied and fixed in time, $n_{\rm dc}$ is the
number of dynamical-core orbitals $n_{\rm dc}$ which are forced to be doubly
occupied but propagated in time, and $n_{\rm act}$ is the number of active orbitals among
which the active electrons are correlated.
We have chosen a laser field having a wavelength of 800 nm with an intensity of 1$\times$10$^{15}$W/cm$^2$. 
We have used a simulation box size of $R_{max}$=300 with $cos^{1/4}$ mask function switched on at 240 
with $L_{max}$=63. 
We have performed a series of calculations starting from 8-active electrons
in 8-active orbitals and gradually increased to 10-active electrons in
14-active orbitals (Fig.~\ref{hhg_active_space}).
We found that the HHG spectrum computed with the configuration (1,0,13)
meet a virtually perfect agreement with the one computed with the
configuration (0,0,14). Therefore, we have chosen the active-space configuration (1,0,13) in
all the following simulations. 
\begin{figure}[hb!]
\centering
\begin{center}
\includegraphics[width=1.1\linewidth]{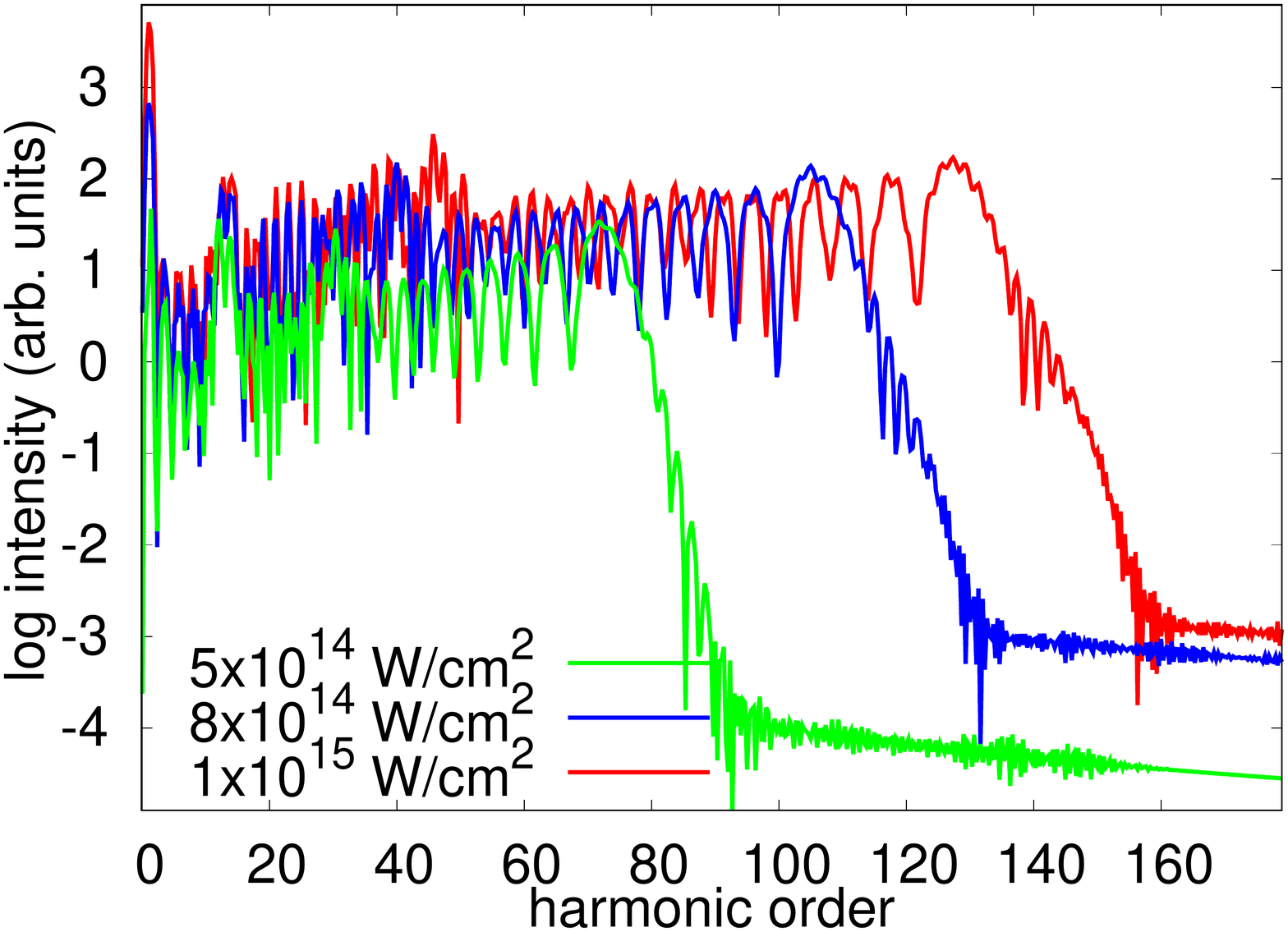}
\caption{HHG spectra of Ne exposed to a laser pulse having a wavelength of 800 nm and varying intensities of 5$\times$10$^{14}$ W/cm$^2$,
8$\times$10$^{14}$ W/cm$^2$, and 1$\times$10$^{15}$ W/cm$^2$, obtained
 with the TD-OMP2 method with an orbital configuration
(1,0,13) and maximum angular momentum $L_{\rm max}$=63.}
\label{hhg_intensity}
\end{center}  
\end{figure}
\begin{figure}[ht!]
\centering
\begin{center}
\includegraphics[width=1.1\linewidth]{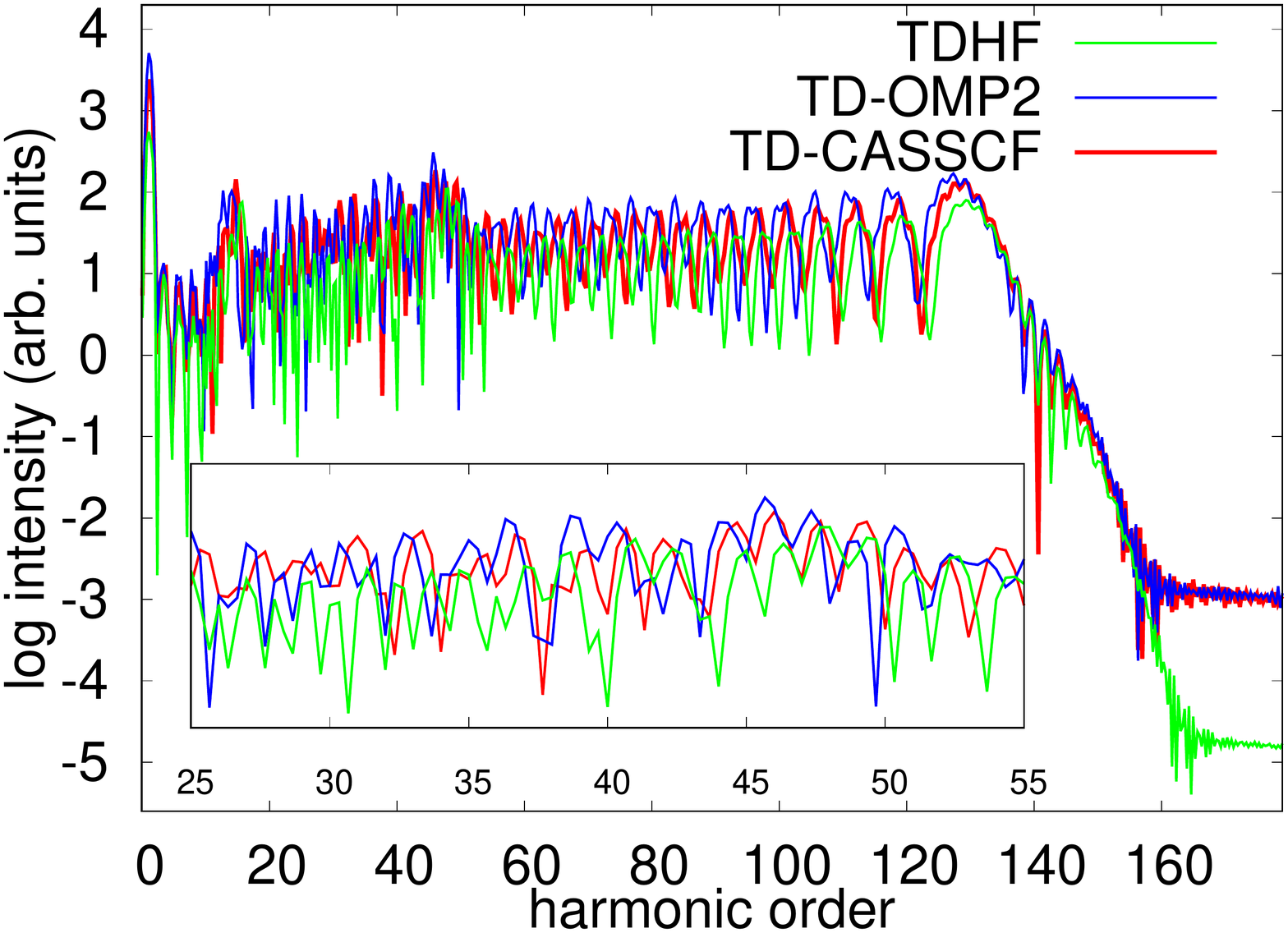}
\caption{HHG spectra of Ne exposed to laser pulse with a wavelength of 800 nm and an intensity 1$\times$10$^{15}$ W/cm$^2$,
Comparison of TD-OMP2 method with TD-CASSCF, and TDHF methods.
Maximum angular momentum $L_{max}$=63 and (1,0, 13) active space configuration for the correlation methods has been used.}
\label{hhg_cas}
\end{center}  
\end{figure}

Next, in Fig.~\ref{hhg_lmax}, we report the convergence pattern of the HHG spectra with respect
to the maximum angular momentum $L_{\rm max}$ to expand the orbitals.
A series of simulations are performed for $51 \leq L_{\rm max} \leq 75$
with an intensity of 1$\times$10$^{15}$ W/cm$^2$ and a wavelength
of 800 nm, with an orbital configuration (1,0,13). As seen in
Fig.~\ref{hhg_lmax} (a), the results steadily tend to converge to the result
obtained employing the highest angular momentum $L_{\rm max}=75$, and
especially, the spectrum with $L_{max}=63$ meets the perfect agreement
with $L_{\rm max}=75$ within the graphical resolution. Therefore, we have chosen
$L_{\rm max}=63$ as the optimum (necessary-and-sufficient) value for the
$L_{\rm max}$ in all the following simulations with an 800 nm wavelength laser
pulse.

Figure~\ref{hhg_intensity} shows comparison of the HHG spectra for the
applied laser intensities 5$\times$10$^{14}$ W/cm$^2$,
8$\times$10$^{14}$ W/cm$^2$, and 1$\times$10$^{15}$ W/cm$^2$ with a
constant wavelength of 800 nm. One can see the dramatic extension of the
cut-off energy with an increase in the applied laser intensity. One also
observes an increase of the harmonic yield by nearly one order of
magnitude in comparison of the lowest and medium intensity cases, and
the saturation of the intensity in comparison of the medium and highest
intensity cases. 
It is important to note that these results are obtained with a
well-calibrated conditions described above, and therefore, reliable as
converged result within the TD-OMP2 approximation.
In Fig. \ref {hhg_cas}, 
we have compared the results from the TD-OMP2 simulations with the fully
correlated TD-CASSCF method, and the uncorrelated TDHF method.

\begin{figure}[ht!]
\centering
\begin{center}
\includegraphics[width=1.1\linewidth]{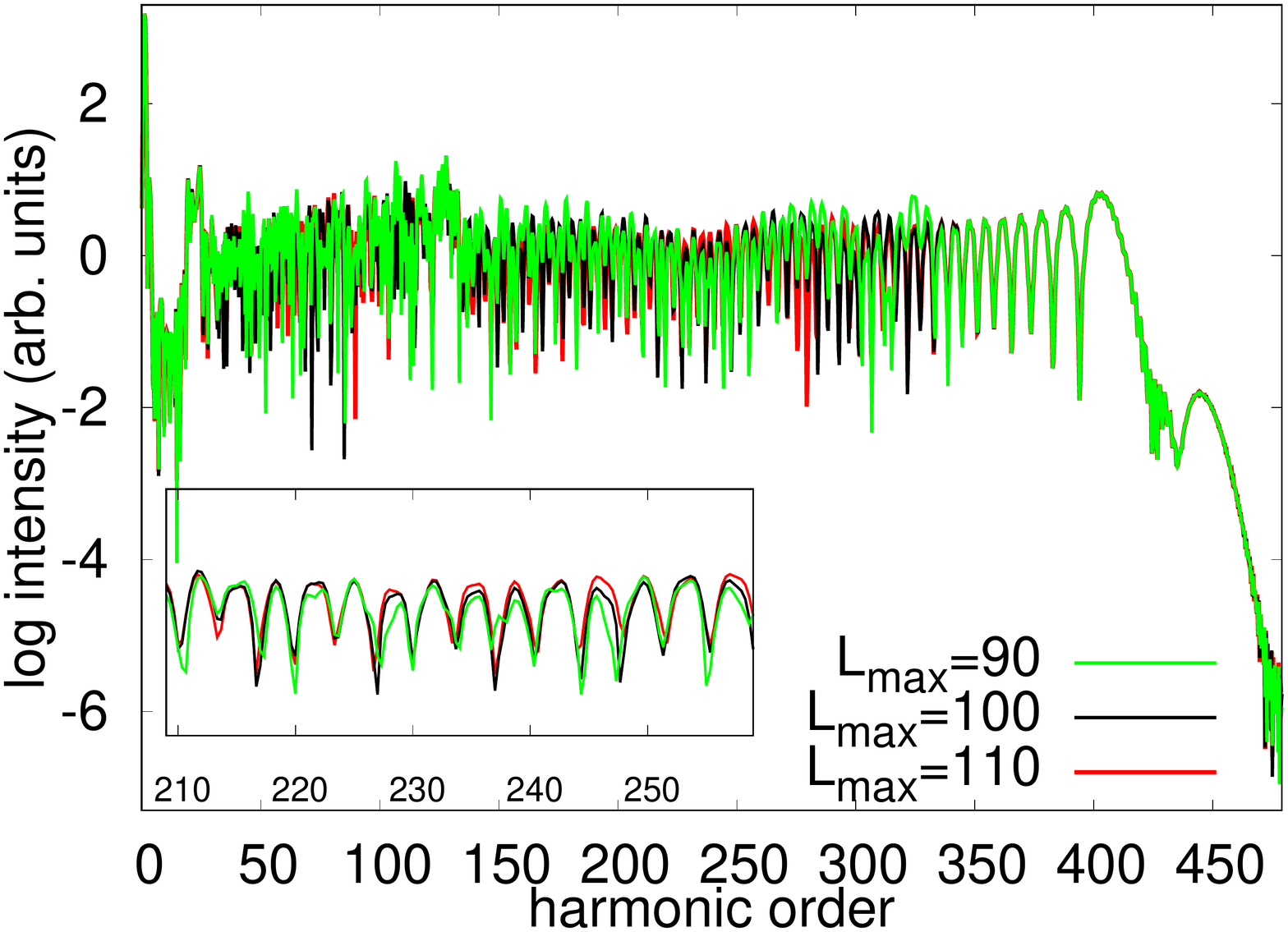}
\caption{HHG spectra of Ne exposed to a laser pulse with a wavelength of 1200 nm and an intensity of 1$\times$10$^{15}$ W/cm$^2$.
Results of the TDHF method obtained with different maximum angular momentum $L_{\rm max}$.}
\label{hhg_lmax_1200}
\end{center}  
\end{figure}
\begin{figure}[ht!]
\centering
\begin{center}
\includegraphics[width=1.1\linewidth]{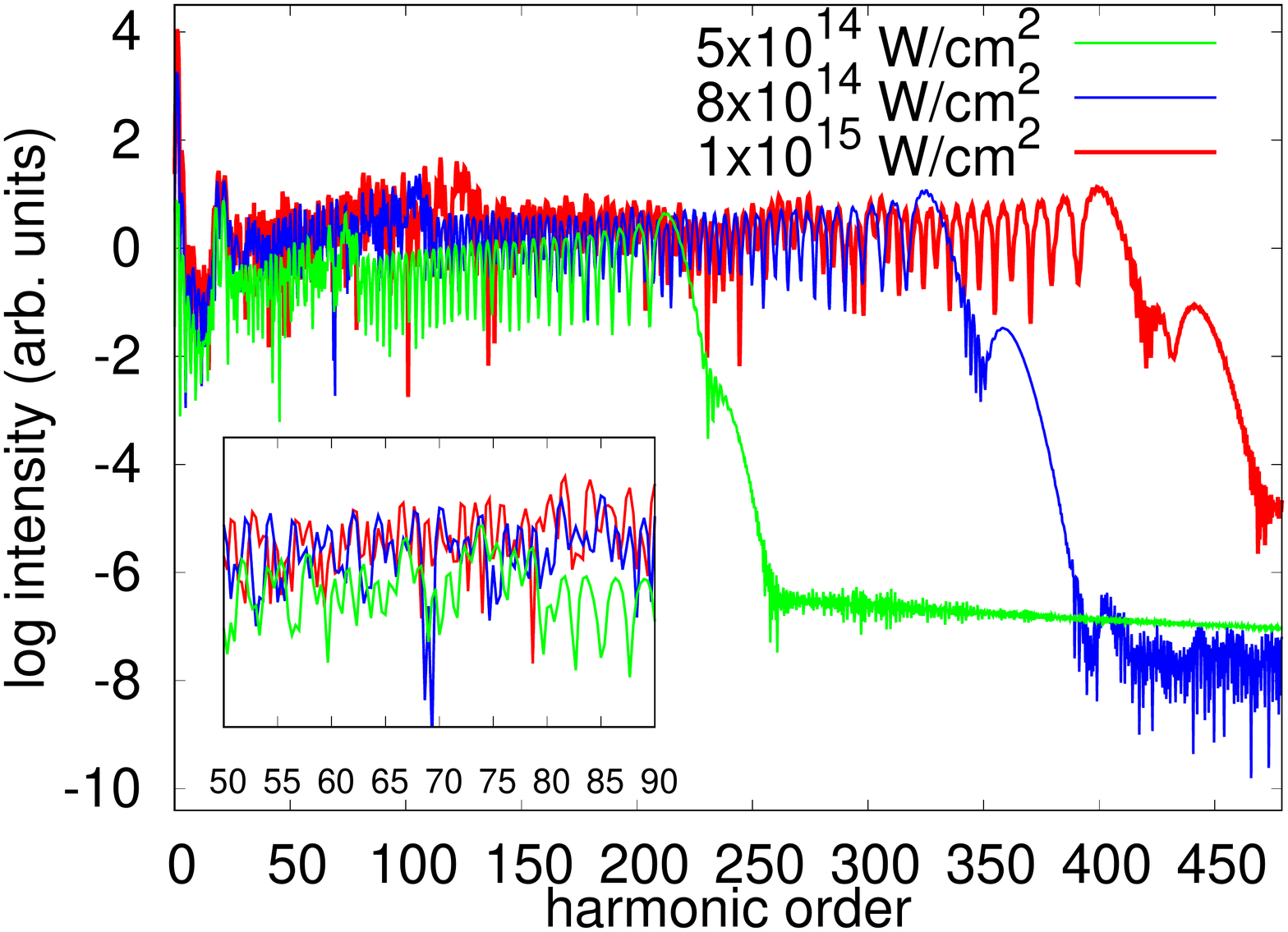}
\caption{HHG spectra of Ne exposed to laser pulse with a wavelength of
 1200 nm and varying intensities of 5$\times$10$^{14}$ W/cm$^2$,
 8$\times$10$^{14}$ W/cm$^2$, and 1$\times$10$^{15}$ W/cm$^2$, obtained
 with TD-OMP2 method with the orbital configuration (1,0,13) and the maximum angular momentum $L_{\rm max}$=100.}
\label{hhg_intensity1}
\end{center}  
\end{figure}
\begin{figure}[ht!]
\centering
\begin{center}
\includegraphics[width=1.1\linewidth]{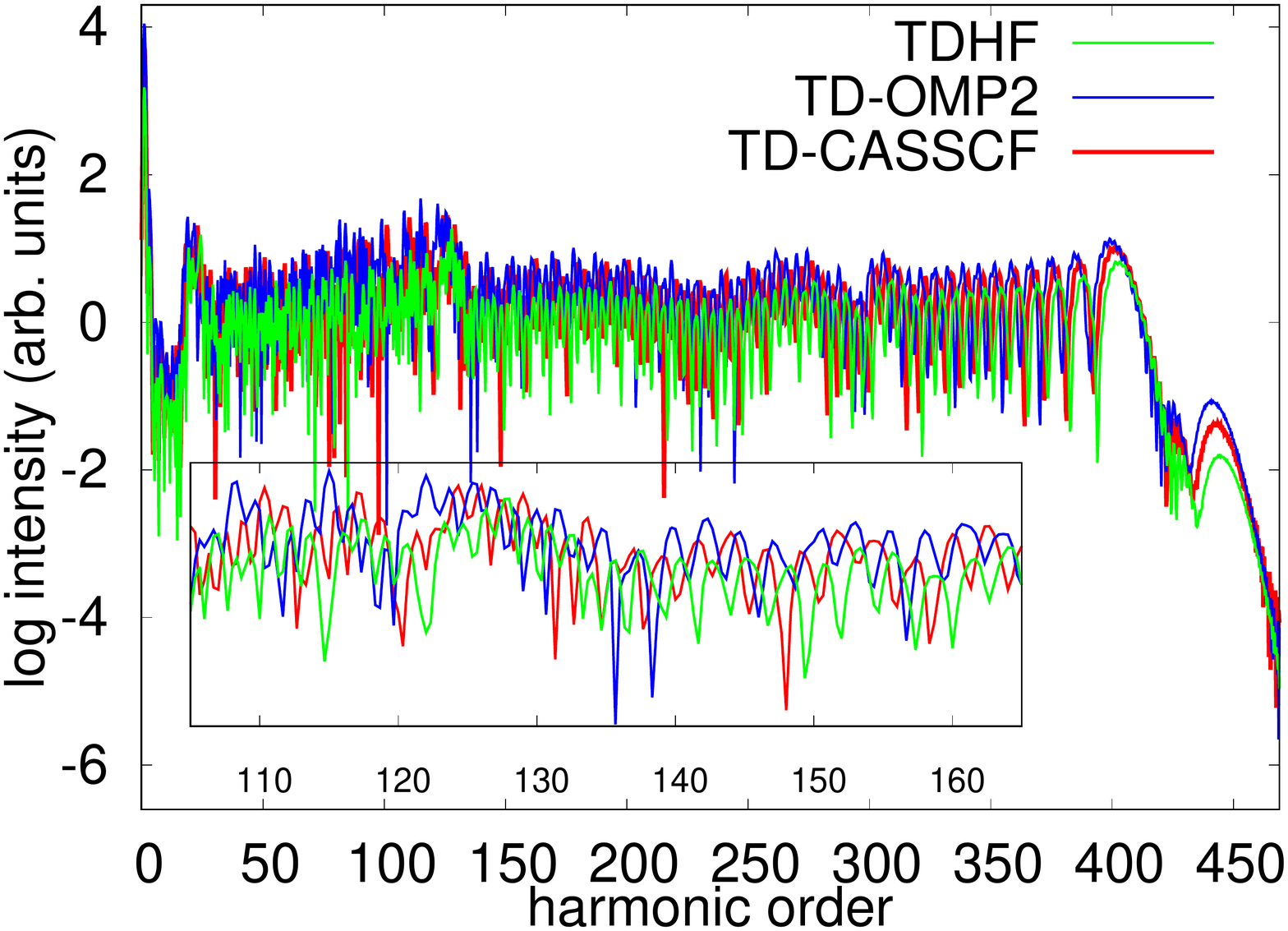}
\caption{HHG spectra of Ne exposed to laser pulse with a wavelength of 1200 nm having intensity of 1$\times$10$^{15}$ W/cm$^2$, Comparison of TD-OMP2 method with TD-CASSCF,
and TDHF method.
Maximum angular momentum $L_{max}$=100 and (1,0, 13) active space configuration for the correlation methods has been used.}
\label{hhg_intensity4}
\end{center}  
\end{figure}


Now we move to a longer wavelength case with $\lambda$ = 1200 nm, which
is a relatively difficult simulation condition, and serves as a robust
test for the newly implemented TD-OMP2 method.
We follow the same procedure as made above for the case with $\lambda =
800$ nm. First we test various $L_{\rm max}$ for the highest-applied intensity of
1$\times$10$^{15}$ W/cm$^2$. We used TDHF method for the calibration
here, and obtained $L_{\rm
max}$=100 as the optimum (necessary-and-sufficient) value for the maximum angular
momentum as confirmed in Fig.~\ref{hhg_lmax_1200}. We, therefore, use
$L_{\rm max}$ = 100 for the following simulations to see the intensity
dependence of the HHG spectra (Fig.~\ref{hhg_intensity1}) obtained with
TD-OMP2, and the comparison of TDHF, TD-OMP2, and TD-CASSCF results
(Fig.~\ref{hhg_intensity4}). 
We could derive essentially the same conclusion from these application
as that for $\lambda$ = 800 nm. The TDHF
severely underestimate the harmonic intensity for the longer wavelength. On the other hand, 
the agreement between the TD-OMP2 and TD-CASSCF spectra is better for
$\lambda$ = 1200 nm than for $\lambda$ = 800 nm. 
Importantly, the agreement between the TD-OMP2 and TD-CASSCF results here
indicates the convergence of the result with respect to the level of
inclusion of the correlation effect, with sufficiently large $L_{\rm
max}$, i.e, at the basis set limit.
In general, TDHF tends to underestimate, and TD-OMP2 overestimates taking TD-CASSCF as the benchmark.
However, the extent of deviation for both TDHF and TD-OMP2 is less for the longer wavelength.

\begin{figure}[ht!]
\centering
\begin{center}
\includegraphics[width=0.9\linewidth]{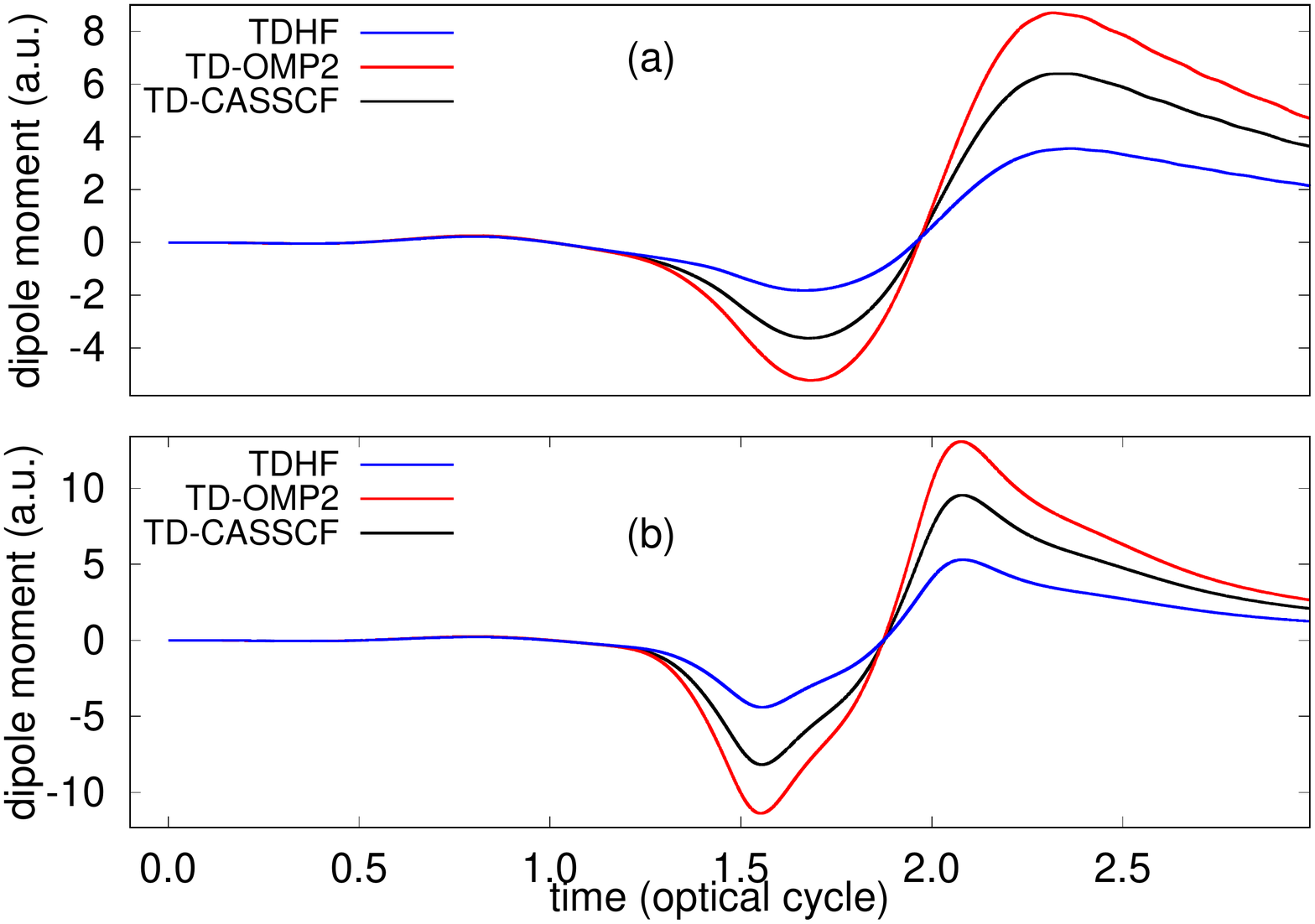}
\caption{Time evolution of the dipole moment of Ne irradiated by
a laser pulse of a wavelength of (a) 800 nm, (b) 1200 nm at an intensity of 1$\times$10$^{15}$W/cm$^2$, calculated
with TDHF, TD-OMP2 and TD-CASSCFcmethods.}
\label{dipole}
\end{center}  
\end{figure}
\begin{figure}[h!]
\centering
\begin{center}
\includegraphics[width=0.9\linewidth]{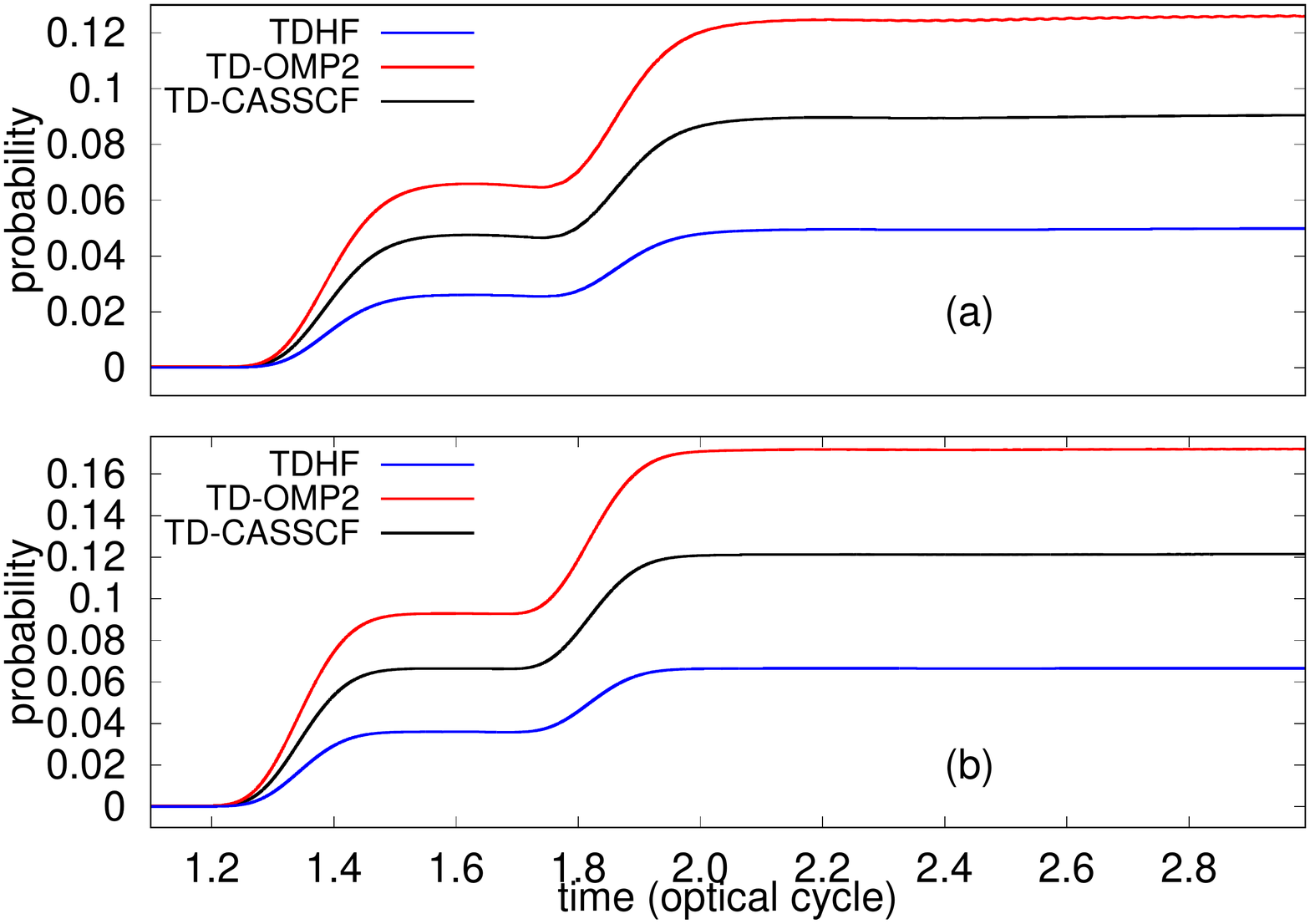}
\caption{Time evolution of single ionization probability of Ne irradiated by
a laser pulse of a wavelength of (a) 800 nm, (b) 1200 nm at an intensity of 1$\times$10$^{15}$W/cm$^2$, calculated
with TDHF, TD-OMP2 and TD-CASSCF methods.}
\label{ion_probability}
\end{center}  
\end{figure}
\begin{figure}[ht!]
\centering
\begin{center}
\includegraphics[width=1.1\linewidth]{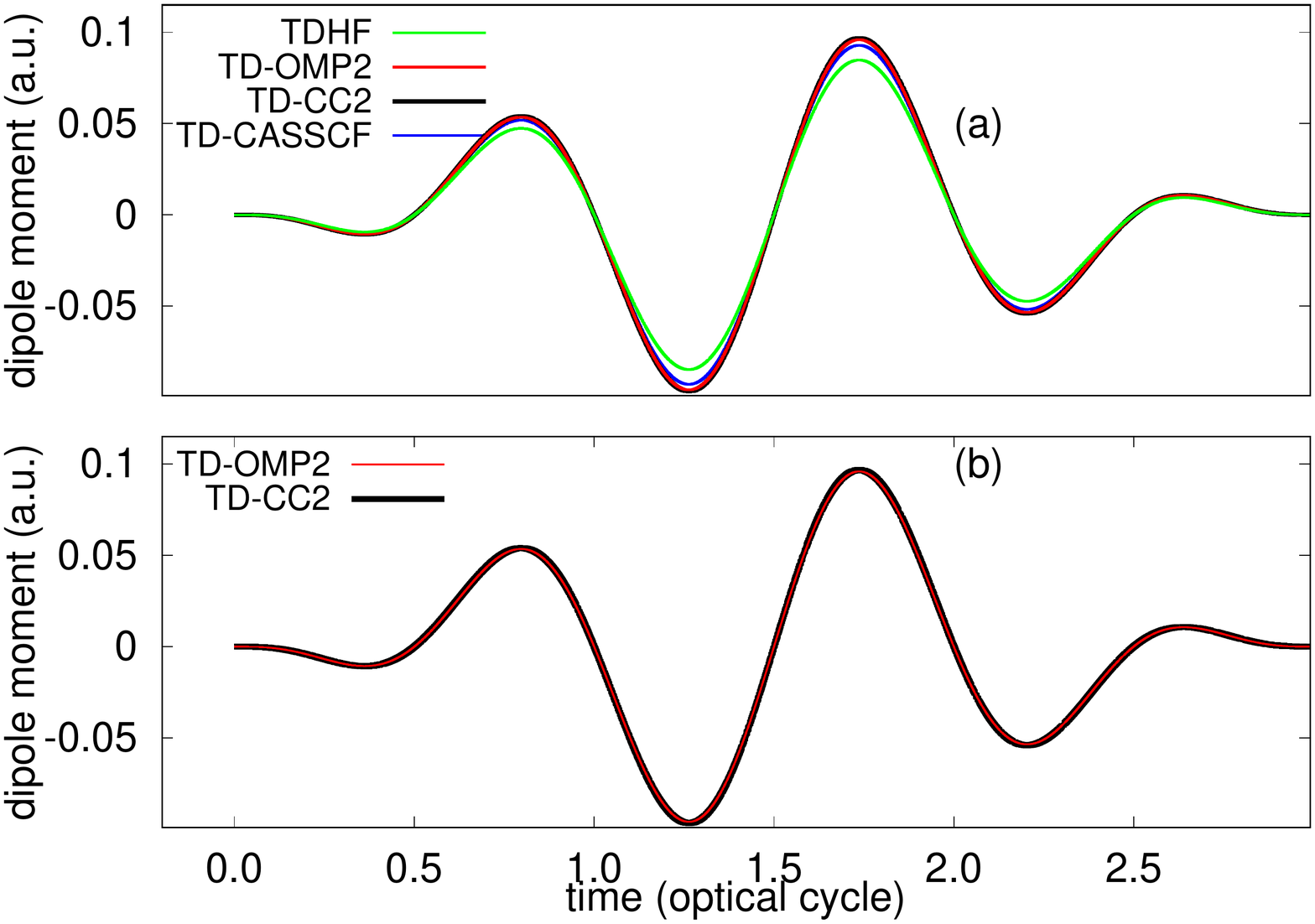}
\caption{Time evolution of the dipole moment of Ne irradiated by
a laser pulse of a wavelength of 800 nm at an intensity of 5$\times$10$^{13}$W/cm$^2$, calculated
with TDHF, TD-OMP2, TD-CC2 and TD-CASSCF methods.}
\label{dipole_cc2}
\end{center}  
\end{figure}
\begin{figure}[ht!]
\centering
\begin{center}
\includegraphics[width=1.1\linewidth]{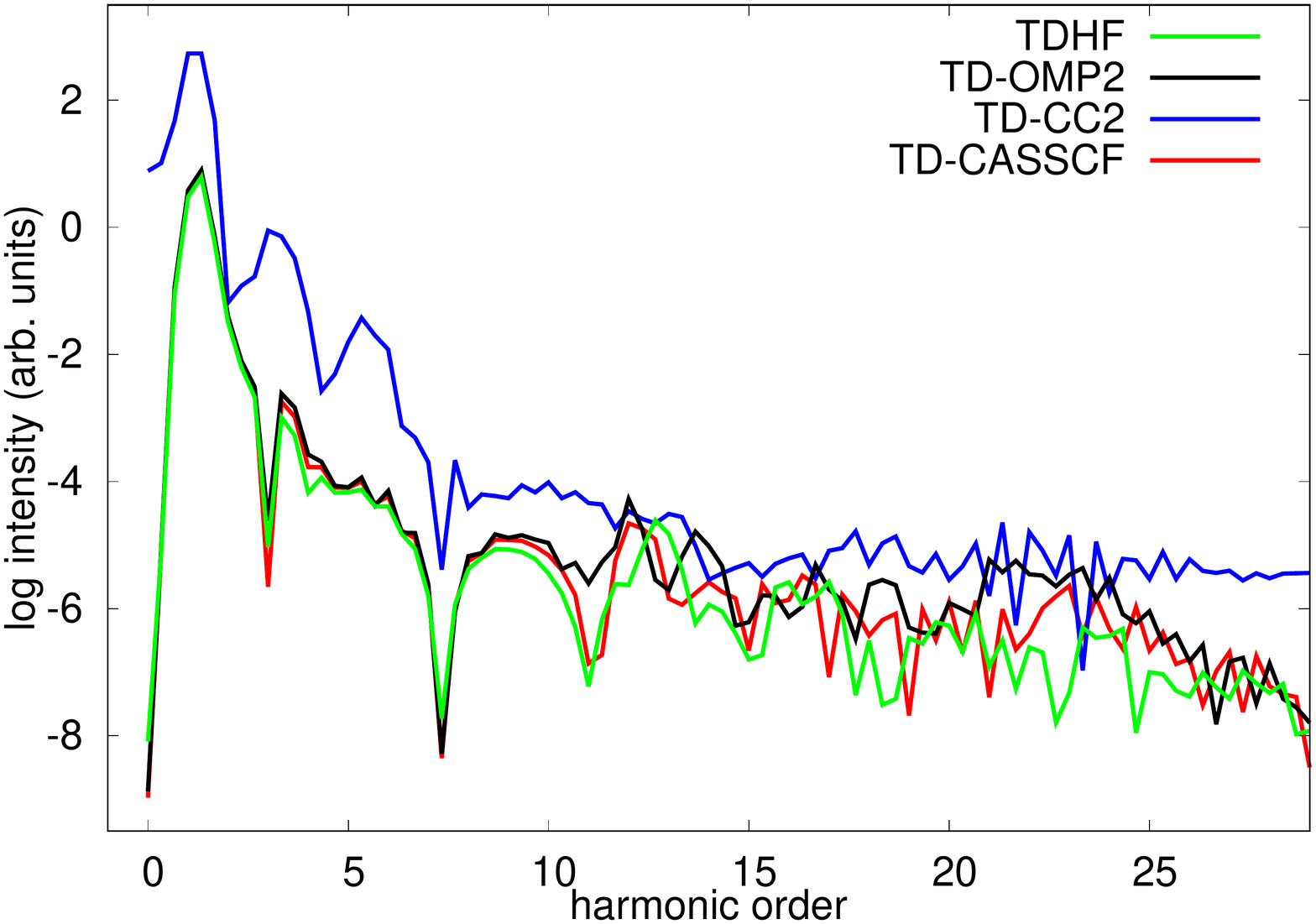}
\caption{HHG spectra of Ne irradiated by
a laser pulse of a wavelength of 800 nm at an intensity of 5$\times$10$^{13}$W/cm$^2$, calculated
with TDHF, TD-OMP2, TD-CC2 and TD-CASSCF methods.}
\label{hhg_cc2}
\end{center}  
\end{figure}


\begin{figure}[ht!]
\centering
\begin{center}
\includegraphics[width=1.1\linewidth]{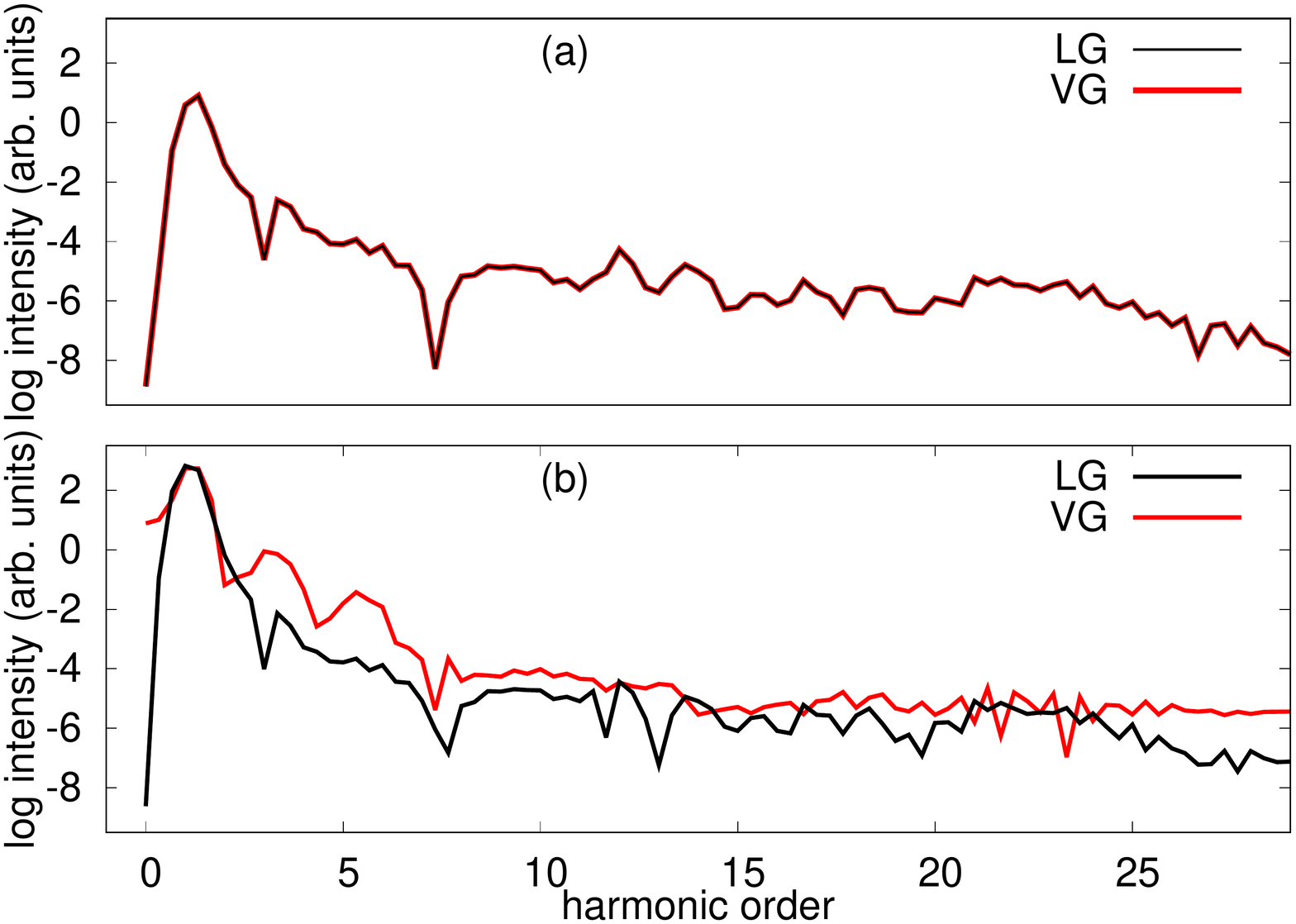}
\caption{HHG spectra of Ne in the length gauge (LG) and velocity gauge (VG) irradiated by
a laser pulse of a wavelength of 800 nm at an intensity of 5$\times$10$^{13}$W/cm$^2$, calculated
with (a) TD-OMP2, and (b) TD-CC2 method.}
\label{hhg_length_velocity}
\end{center}  
\end{figure}

To establish our findings further, we report the time-evolution of dipole moment and single ionization probability
at an intensity of 1$\times$10$^{15}$W/cm$^2$ having a wavelength of (a) 800 nm,
and (b) 1200 nm in Fig. ~\ref{dipole} and Fig.~\ref{ion_probability}, respectively.
The dipole moment is calculated as a trace ${\langle \psi_p|\hat z|\psi_q \rangle\rho_p^q}$, and
the single ionization probability is evaluated as the probability of finding an electron outside a sphere of radius 20 a.u.
We have used the same optimized simulation condition for each wavelength, as reported earlier.
In general, TDHF tends to underestimate, whereas TD-OMP2 overestimates in comparison to the TD-CASSCF; the deviation is more for TDHF, however.
Such a convergence pattern for these methods is often encountered in the ground state calculations also.
The difference from the TD-CASSCF for both TDHF and TD-OMP2 reduces for the longer wavelength (Fig. \ref{dipole}(b)), however, for the ionization probability,
the difference between  these methods remains nearly constant shown in Fig.\ref{ion_probability} (b).
The role of electron correlation is less 
for the longer wavelength, as it is the outer valence electrons, which are driven by the incident laser.

In Fig.~\ref{dipole_cc2}, we have compared time-evolution of dipole-moment of Ne with all these methods at an intensity of 5$\times$10$^{13}$W/cm$^2$ at a wavelength
of 800 nm.
In these simulations, we also correlated 8 electrons in 13 active orbitals, and laser electric field taken in the velocity gauge.
We are unsuccessful in obtaining a stable convergence for the TD-OCC2 method using laser intensities employed for the TD-OMP2 simulations.
The ionization potential of Ne is high, and the employed intensity is low enough to have sufficient ionization.
Therefore, the role of electron correlation is not that relevant in this particular case, and all the methods predict nearly identical time-dependent dipole moment.
In Fig.~\ref{hhg_cc2}, we have reported HHG spectra.
Except for the TD-CC2 method, all the other methods produce similar spectra.
It overestimates high-harmonic intensity from all other methods. The outcome makes a clear case for the need of a low-scaling orbital-optimized theory.
In Fig.~\ref{hhg_length_velocity}, we compare HHG spectra taking the laser-electric 
field in the length gauge with the result obtained with the velocity gauge treatment using the TD-OMP2 (Fig. a) and TD-CC2 (Fig. b) method.
We have used identical simulations conditions ($l_{max}, r_{max}$, active space configuration) for these length gauge simulations as used for all other velocity gauge treatment using 
800 nm wavelength laser.
The outcome suggests that the TD-CC2 method does not provide a gauge-invariant description of properties of interest, whereas TD-OMP2 does, which make
TD-OMP2 a superior choice for the study of strong-field dynamics.

In the TD-OMP2 method, the doubly excited determinants are approximately incorporated in the configuration space. Therefore,
it takes into account at least a part of the electron correlation, which makes it overall a better performer in comparison to the TDHF, 
where the electron correlation is missing.
The electron correlation is more important in the far part of the spectrum, where the intensity profile of the spectrum drops quickly for TDHF,
or in other words, the difference with the TD-CASSCF is more for the TDHF than TD-OMP2.

\begin{table}[!h]
\caption{\label{tab:timing_electron} \color{black} Comparison of the CPU
time (in second) spent for the evaluation of the $T_1$, $\Lambda_1$, $T_2$,
$\Lambda_2$ equation, 1RDM, and 2RDM for TD-CC2 and TD-OMP2 methods.
{\color{black} CPU time spent for the simulation of Ne atom for   
 1000 time steps ($0 \leq t \leq 0.1T$) of a real-time simulation
 ($I_0=5\times 10^{13}$ W/cm$^{2}$ and $\lambda=800$ nm),
 using an Intel(R) Xeon(R) Gold 6230 CPU with 40 processors having a clock speed of 2.10GHz.}
}
\begin{center}
\begin{tabular}{rrrrrrrrrrrrrr}
\hline
\multicolumn{6}{c}{TD-CC2} & & & \multicolumn{6}{c}{TD-OMP2}\\
\cline{1-6} \cline{8-14}\\
\multicolumn{1}{c}{T$_1$}&\multicolumn{1}{c}{$\Lambda_1$}&\multicolumn{1}{c}{T$_2$}&\multicolumn{1}{c}{$\Lambda_2$}&\multicolumn{1}{c}{1RDM}& \multicolumn{1}{c}{2RDM}
&&&\multicolumn{1}{c}{T$_1$}&\multicolumn{1}{c}{$\Lambda_1$}&\multicolumn{1}{c}{T$_2$}&\multicolumn{1}{c}{$\Lambda_2$}&\multicolumn{1}{c}{1RDM}& \multicolumn{1}{c}{2RDM}
\\
\hline
\hline \\
1.31&\,10.47&\,8.59&\,2.47&\,2.48&\,17.46 &&&\,- &\,- &\, 0.71&\,- &\, 1.06 &\, 0.67\\
\hline
\hline
\end{tabular}
\end{center}
%
\end{table}
In Tab. \ref{tab:timing_electron}, we compare computational timing for 1000 time-step propagation in various parts of TD-CC2 and TD-OMP2
methods for the real-time simulations.
Even though the overall computational scaling for both the methods is the same ($N^5$), the TD-OMP2 does not involve solutions of the $T_1$, $\Lambda_1$, and the $\Lambda_2$
amplitude equations.
The $\Lambda_2$ amplitudes are complex conjugate of the $T_2$ amplitudes due to the linear structure of the functional.
The $T_2$ equation for the TD-CC2 method has many terms and involves multiple operator products.
The time saving for the TD-OMP2 method comes from the evaluation of 2RDMs, which scales N$^4$ and does not involve any operator products.
On the other hand, it is $N^5$ for the TD-CC2 method.

\section{Concluding remarks}\label{sec4}
In this article, we have applied the recently developed TD-OMP2 method to compute the HHG spectra of Ne atom as a case study to analyze
the performance of the implemented method in demanding laser conditions.
Further, we have implemented the TD-CC2 method, which is also an N$^5$ scaling second-order approximation to the parent TD-CCSD method.
The TD-CC2 method does not provide a gauge-invariant description of the properties of interest and is not stable with rigorous simulations conditions,
often required while studying strong-field dynamics.
On the contrary, TD-OMP2 is very stable, does not breakdown even with harsher simulation conditions, and it is gauge-invariant.
Additionally, TD-OMP2 is computationally more favorable.
All these make TD-OMP2 as a superior choice over the TD-CC2 method.
While the performance of the TD-OMP2 method is moderate,
it is remarkable that such highly nonlinear nonperturbative phenomena
can be stably computed within the
framework of time-dependent perturbation method, by virtue of the nonperturbative inclusion of the laser-electron interacdtion and time-dependent optimization of orbitals.
This will open a way to correlated time-dependent calculation for large chemical systems,
for which the applications of multiconfiguration methods such as TD-CASSCF are challenging.
The method will also be useful to study moderate size systems;
calibrations of simulation conditions and extensive parameter surveys
can be performed before stepping into simulations with more rigorous and therefore, more expensive method.

\section*{Acknowledgements}
This research was supported in part by a Grant-in-Aid for
Scientific Research (Grants No. 16H03881, No. 17K05070,
No. 18H03891, and No. 19H00869) from the Ministry of Education, Culture,
Sports, Science and Technology (MEXT) of Japan. 
This research was also partially supported by JST COI (Grant No.~JPMJCE1313), JST CREST (Grant No.~JPMJCR15N1),
and by MEXT Quantum Leap Flagship Program (MEXT Q-LEAP) Grant Number JPMXS0118067246.
This research was also partially supported by Startup funding of Graduate School of Engineering, The University of Tokyo.

\section*{Disclosure statement}
No potential conflict of interest was reported by the authors.
\section*{Funding}
This research was supported in part by a Grant-in-Aid for
Scientific Research (Grants No. 16H03881, No. 17K05070,
No. 18H03891, and No. 19H00869) from the Ministry of Education, Culture,
Sports, Science and Technology (MEXT) of Japan. 
This research was also partially supported by JST COI (Grant No.~JPMJCE1313), JST CREST (Grant No.~JPMJCR15N1),
and by MEXT Quantum Leap Flagship Program (MEXT Q-LEAP) Grant Number JPMXS0118067246.


\appendix
\section{Algebraic details of TD-CC2}\label{app:td-cc2}
The EOMs for the amplitudes are given by
\begin{eqnarray}
i\dot \tau_i^a&=& f_i^a+ f_c^k\tau_{ki}^{ca}+0.5 v_{cd}^{ak}\tau_{ik}^{cd}-0.5 v_{ic}^{kl}\tau_{kl}^{ac}+ f_c^a\tau_i^c- f_i^k\tau_k^a+v_{ic}^{ak}\tau_k^c\nonumber\\
            &&+v_{cd}^{kl}\tau_{kl}^{ad}\tau_i^c-0.5v_{cd}^{kl}\tau_{il}^{cd}\tau_k^a+v_{cd}^{kl}\tau_l^d\tau_{ki}^{ca}- f_c^k\tau_i^c\tau_k^a+v_{cd}^{ak}\tau_k^d\tau_i^c\nonumber\\
            &&-v_{ic}^{kl}\tau_l^c\tau_k^a-v_{cd}^{kl}\tau_l^d\tau_i^c\tau_k^a\label{t1matrix}
\end{eqnarray}

\begin{eqnarray}
i\dot \tau_{ij}^{ab}&=&v_{ij}^{ab}+p(ab) f_c^a \tau_{ij}^{ab}-p(ij) f_i^k \tau_{kj}^{ab}+p(ij)v_{ic}^{ab}\tau_j^c-p(ab)v_{ij}^{ak}\tau_k^b-p(ij) f_c^k\tau_i^c\tau_{kj}^{ab}\nonumber\\ 
                  &&-p(ab) f_c^k\tau_k^a\tau_{ij}^{cb}+v_{cd}^{ab}\tau_i^c\tau_j^d+v_{ij}^{kl}\tau_k^a\tau_l^b-p(ij|ab)v_{ic}^{ak}\tau_k^b\tau_j^c-p(ij|ab)v_{cd}^{ka}\tau_i^d\tau_j^c\tau_k^b\nonumber \\
                  &&+p(ij)v_{cj}^{kl}\tau_k^a\tau_i^c\tau_l^b+v_{cd}^{kl}\tau_i^c\tau_k^a\tau_j^d\tau_l^b\label{t2matrix}
\end{eqnarray}

\begin{eqnarray}
-i\dot \lambda_a^i&=&f_a^i+v_{ab}^{ij}\tau_j^b+ f_a^b\lambda_b^i- f_j^i\lambda_a^j+v_{aj}^{ib}\lambda_b^j- f_c^i \tau_k^c\lambda_a^k- f_a^k \tau_k^b\lambda_b^i\nonumber\\
                  &&+v_{ab}^{cj}\tau_j^b\lambda_c^i-v_{kb}^{ij}\tau_j^b\lambda_a^k+v_{ac}^{ib}\tau_j^c\lambda_b^j-v_{ak}^{il}\tau_l^b\lambda_b^k+v_{ac}^{il}\tau_{lj}^{cb}\lambda_b^j\nonumber\\
                  &&-0.5 v_{cb}^{ij}\tau_{jk}^{bc}\lambda_a^k-0.5v_{ac}^{kj}\tau_{kj}^{bc}\lambda_b^i-v_{ab}^{ij}\tau_j^c\tau_k^b\lambda_c^k-v_{cb}^{ij}\tau_j^b\tau_k^c\lambda_a^k\nonumber\\
                  &&-v_{ab}^{kj}\tau_j^b\tau_k^c\lambda_c^i+0.5 v_{aj}^{cb}\lambda_{cb}^{ij}-0.5 v_{kj}^{ib}\lambda_{ab}^{kj}+0.5 v_{ad}^{cb}\tau_j^d\lambda_{cb}^{ij}\nonumber\\
                  &&+0.5v_{kl}^{ij}\tau_j^b\lambda_{ab}^{kl}-v_{jc}^{bi}\tau_l^c\lambda_{ab}^{lj}-v_{ja}^{bk}\tau_k^c\lambda_{bc}^{ji}-0.5 f_c^i\tau_{kj}^{cb}\lambda_{ab}^{kj}\nonumber\\
                  &&- f_a^k\tau_{kj}^{cb}\lambda_{cb}^{ij}-v_{ab}^{dj}\tau_j^c\tau_k^b\lambda_{dc}^{ik}+v_{lb}^{ij}\tau_k^b\tau_j^c\lambda_{ac}^{lk}-0.5v_{bc}^{id}\tau_k^c\tau_j^b\lambda_{da}^{kj}\nonumber\\
                  &&+0.5 v_{al}^{kj}\tau_j^b\tau_k^c\lambda_{cb}^{il}+0.5v_{bd}^{ik}\tau_l^d\tau_k^c\tau_j^b\lambda_{ac}^{jl}+0.5v_{ac}^{jk}\tau_k^d\tau_l^c\tau_j^b\lambda_{bd}^{il}\label{l1matrix}
\end{eqnarray}

\begin{eqnarray}
-i\dot \lambda_{ab}^{ij}&=&v_{ab}^{ij}+p(ij|ab)f_a^i\lambda_b^j+p(ij|ab)v_{ac}^{ik}\tau_k^c\lambda_b^j+p(ij)v_{ab}^{ic}\lambda_c^j\nonumber\\
                        &&-p(ab)v_{ak}^{ij}\lambda_b^k-p(ij)v_{ab}^{ik}\tau_k^c\lambda_c^j-p(ab)v_{ac}^{ij}\tau_k^c\lambda_b^k+p(ab)f_a^c\lambda_{cb}^{ij}\nonumber\\
                        &&-p(ij) f_k^i\lambda_{ab}^{kj}-p(ij) f_c^i\tau_k^c\lambda_{ab}^{jk}-p(ab) f_a^k \tau_k^c\lambda_{cb}^{ij}\label{l2matrix}
\end{eqnarray}

The correlation contributions to the RDMs for the CC2 method is given by
\begin{eqnarray}
&&\gamma_a^i=\lambda_a^i, \gamma_i^a=\tau_i^a+\lambda_b^j\tau_{ij}^{ab}+\lambda_b^j\tau_i^b\tau_j^a-0.5\lambda_{cd}^{kl}\tau_{ki}^{cd}\tau_l^a-0.5 \lambda_{cd}^{kl}\tau_{kl}^{ca}\tau_i^d\nonumber \\
&&\gamma_j^i=\lambda_a^i\tau_j^a-0.5 \lambda_{cb}^{ki}\tau_{kj}^{cb}, \gamma_b^a=\lambda_b^i\tau_i^a+0.5\lambda_{cb}^{kl}\tau_{kl}^{ca}\label{cc2_1rdm}
\end{eqnarray}
\begin{eqnarray}
&&\Gamma_{cd}^{ab}=\lambda_{cd}^{ij}\tau_i^a\tau_j^b, \Gamma_{kl}^{ij}=\lambda_{cd}^{ij}\tau_k^c\tau_l^d, 
\Gamma_{ak}^{bc}=\lambda_a^l\tau_{lk}^{bc}+p(bc)\lambda_a^l\tau_l^b\tau_k^c,
\Gamma_{jk}^{ic}=-\lambda_b^i\tau_{jk}^{bc}-p(jk)\lambda_b^i\tau_j^b\tau_k^c\,\,\,\,\,\,\,\,\,\,\,\,\,\,\,\,\nonumber\\
&&\Gamma_{al}^{bc}=-\lambda_{ad}^{ij}\tau_i^a\tau_j^c\tau_l^d, \Gamma_{jl}^{ic}=\lambda_{ad}^{ik}\tau_j^a\tau_k^c\tau_l^d, \Gamma_{ca}^{jb}=\lambda_{ca}^{ji}\tau_i^b, \Gamma_{ci}^{ka}=-\lambda_{ac}^{ik}\tau_i^a\nonumber\\
&&\Gamma_{ij}^{ab}=\tau_{ij}^{ab}+0.5p(ij|ab)\tau_i^a\tau_j^b+p(ij|ab)\lambda_c^k\tau_{ki}^{ca}\tau_j^b-p(ij)\lambda_c^k\tau_i^c\tau_{kj}^{ab}-p(ab)\lambda_c^k\tau_k^a\tau_{ij}^{cb}\nonumber\\
&&-p(ij|ab)\lambda_c^k\tau_k^a\tau_i^c\tau_j^b+\lambda_{cd}^{kl}\tau_i^c\tau_j^d\tau_k^a\tau_l^b\nonumber \\
&&\Gamma_{aj}^{ib}=\lambda_a^i\tau_j^b-\lambda_{ac}^{ik}\tau_k^b\tau_j^c, \Gamma_{ab}^{ij}=\lambda_{ab}^{ij}\label{cc2_2rdm}
\end{eqnarray}

\appendix
\section{Ground-state energy}
\begin{table}[!h]
\caption{\label{tab:comparison} Comparison of ground state energies of Be, and BH (r$_e$=2.4 bohr, within a (6, 6) active space configuration).
The overlap, one-electron, and two-electron repulsion
integrals over Gaussian basis functions are obtained from  
Gaussian09 program (Ref.~\citenum{gaussian09}).
Imaginary time relaxation is used to obtain the ground in the orthonormalized Gaussian basis.
A convergence cutoff of 10$^{-15}$ Hartree of energy difference is used in the subsequent time-step.}
\begin{tabular}{lllcc}
\hline
&
Basis &
Method&
\multicolumn{1}{c}{This work} &
\multicolumn{1}{c}{Reference} \\
\hline
\hline
Be & cc-pVDZ\cite{ne_basis}   &HF    &$-$14.5723\,\,376 & $-$14.5723\,\,376\cite{psi4}\\
   &                          &MP2   &$-$14.5986\,\,736 & $-$14.5986\,\,736\cite{psi4}\\
   &                         &OMP2   &$-$14.5987\,\,486 & $-$14.5987\,\,486\cite{psi4}\\
   &                          &CC2  &$-$14.5988\,\,233 &$-$14.5988\,\,233\cite{psi4}\\
   &                          &FCI   &$-$14.6174\,\,095 & $-$14.6174\,\,095\cite{psi4}\\ 
\hline
BH & DZP\cite{harrison1983full}&HF    &$-$25.1247\,\,420& $-$25.1247\,\,42\cite{krylov1998size}\\
   &                         &MP2   &$-$25.1325\,\,603& \\
   &                         &OMP2 &$-$25.1528\,\,754&\\
   &                         &CC2  &$-$25.1325\,\,787& \\
   &                         &$^a$OCC2 &$-$25.1528\,\,704&                    \\
   &                         &CASSCF&$-$25.1783\,\,349& $-$25.1783\,\,35\cite{krylov1998size}\\
\hline   
\hline
\end{tabular}\\
$^a$It does not include hole-particle rotations.
\end{table}
To check the correctness of the newly implemented TD-CC2 method, we have done a series of calculations taking Be and BH as example systems.
We have assessed the correctness of the implementation of the TD-OMP2 in an earlier article \cite{pathak2020mp2}.
The required one-electron, two-electron, and overlap matrix elements are obtained from the Gaussian09 \cite{gaussian09} and orthonormalized
to interface with our numerical code.
In these calculations, we have taken the number of grid points as the same as the number of Gaussian basis functions for a chosen basis set.
For Be, we have used cc-pVDZ \cite{ne_basis} basis set, and  all the orbitals are taken as active to compare with the PSI4 \cite{psi4}, which 
only supports all orbitals as active.
Our implementation allows a flexible classification of the orbital subspace into frozen-core, dynamical core, and active, however.
We have used DZP basis \cite{harrison1983full} for BH. All 6 electrons are chosen as active 
and distributed among 6 active orbitals for the correlation methods to check the correctness for the active space implementation. 
The OCC2 method does not include orbital rotation among hole particle subspace, which encounters convergence difficulty while 
retaining single excitation amplitudes \cite{scuseria1987optimization}.
We have tabulated our results against the values obtained by Krylov {\it et. al,} \cite{krylov1998size} in Tab.~\ref{tab:comparison}.
Our results are identical to the available values.

\end{document}